\documentclass[%
 reprint,
 amsmath,amssymb,
 aps,
]{revtex4-2}

\usepackage{graphicx}
\usepackage{dcolumn}
\usepackage{bm}
\usepackage[colorlinks=true,urlcolor=blue,citecolor=black,linkcolor=black]{hyperref}
\usepackage{url}
\begin{document}

\preprint{APS/123-QED}

\title{Non-Gaussian reconciliation for continuous-variable quantum key distribution}
\author{Xiangyu Wang$^1$}
\email{xywang@bupt.edu.cn}
\author{Menghao Xu$^1$}
\author{Yin Zhao$^1$}
\author{Ziyang Chen$^2$}
\email{chenziyang@pku.edu.cn}
\author{Song Yu$^1$}
\author{Hong Guo$^2$}

\affiliation{$^1$State Key Laboratory of Information Photonics and Optical Communications, Beijing University of Posts and Telecommunications, Beijing 100876, China}

\affiliation{$^2$State Key Laboratory of Advanced Optical Communication Systems and Networks, School of Electronics and Center for Quantum Information Technology, Peking University, Beijing 100871, China}

\date{\today}

\begin{abstract}
Non-Gaussian modulation can improve the  performance of continuous-variable quantum key distribution (CV-QKD). For Gaussian modulated coherent state CV-QKD, photon subtraction can realize non-Gaussian modulation, which can be equivalently implemented by non-Gaussian postselection. However, non-Gaussian reconciliation has not been deeply researched, which is one of the key technologies in CV-QKD. In this paper, we propose a non-Gaussian reconciliation method to obtain identical keys from non-Gaussian data. Multidimensional reconciliation and multi-edge type low density parity check codes (MET-LDPC) are used in non-Gaussian reconciliation scheme, where the layered belief propagation decoding algorithm of MET-LDPC codes is used to reduce the decoding complexity. Furthermore, we compare the error correction performance of Gaussian data and non-Gaussian data. The results show that the error correction performance of non-Gaussian data is better than Gaussian data, where the frame error rate can be reduced by 50\% for code rate 0.1 at SNR of 0.1554 and the average iteration number can be reduced by 25\%.
\end{abstract}

\maketitle

\section{INTRODUCTION}
Secure communication needs secret keys. However, the classical key generation algorithm based on computational complexity is seriously threatened by quantum computer and new mathematical algorithm. In order to solve this problem, scholars have proposed quantum key distribution (QKD) protocols based on the basic principles of quantum physics, which is one of the most mature quantum information technology~\cite{gisin2002quantum,xu2020secure,pirandola2020advances}. QKD allows two separate parties (Alice and Bob) to establish unconditional secure keys through an unsecure quantum channel which maybe controlled by potential eavesdroppers (Eve). According to the encoding dimension of quantum states, QKD is divided into two branches, discrete-variable (DV) QKD~\cite{bennett1984quantum,wang2022twin} and continuous-variable (CV) QKD~\cite{grosshans2002continuous,grosshans2003quantum,weedbrook2012gaussian,jouguet2013experimental}. These two kinds of protocols both have unconditional security~\cite{garcia2006unconditional,furrer2012continuous,leverrier2015composable,jain2022practical}, in which Gaussian modulated coherent states CV-QKD protocols have the advantages of being compatible with classical coherent optical communication technology~\cite{chen2023continuous,wang2020high}. However, due to the imperfection of the practical system devices and the reconciliation efficiency of postprocessing, the transmission distance of CV-QKD system is limited~\cite{shen2021strengthening,wang2017efficient,wang2019realistic}. Therefore, improving the secret key rate at given distance~\cite{huang2015continuous,wang2018high} and extending the transmission distance of the system are two improtant development trends in the field of CV-QKD~\cite{jouguet2013experimental,huang2016long,wang2018long}.

Recently, CV-QKD has made great progress in theory~\cite{ghorai2019asymptotic,lin2019asymptotic,huang2021realizing} and experiment~\cite{tian2022experimental,wang2022sub}. The transmission distance has been significantly improved due to the optimization of experiment setups and the improvement of reconciliation efficiency~\cite{huang2016long,zhou2019continuous}. Some new protocols have been proposed to improve the performance of CV-QKD system, such as the noiseless linear amplification~\cite{blandino2012improving} and photon subtraction~\cite{opatrny2000improvement,huang2013performance}. These two quantum operation can extend the transmission distance. However, due to the imperfection of actual devices and other factors, it is difficult to implement in physical system. Thus, it is hard to achieve the ideal effect. To solve these problems, some postselection protocols are proposed. Through Gaussian postselection, virtual noiseless linear amplification can be realized~\cite{fiuravsek2012gaussian}, it has been demonstrated experimentally~\cite{chrzanowski2014measurement}. The physical realization of photon subtraction operation requires an ideal single photon detector. Therefore, the implementation cost will be increased, the effect will be reduced due to the imperfection of the actual devices. Non-Gaussian postselection is proposed to implement virtual photon subtraction, which avoids the use of single photon detector and complex physical operation~\cite{li2016non}. The raw data after Gaussian postselection is still following Gaussian distribution, while for the virtual photon subtraction inside Alice the raw data of Alice after non-Gaussian postselection is no longer following Gaussian distribution.

\begin{figure*}[htb]
	\centering
	\includegraphics[width=\linewidth]{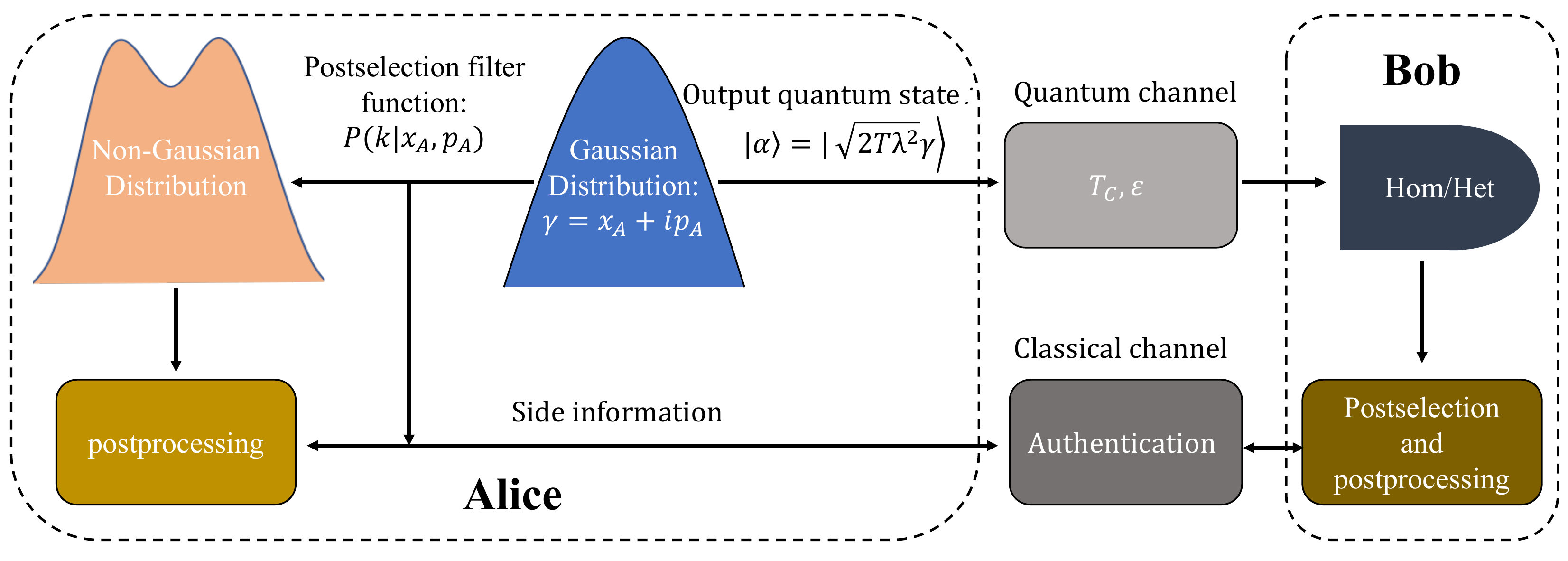}
	\caption{The PM scheme of CV-QKD with non-Gaussian postselection in Alice. $P(k|x_A,p_A)$ is the postselection filter function. $\gamma$ are Alice's raw data. $\left| \alpha  \right\rangle$ is output quantum state. $T_C$ is the transmittance of quantum channel. $\epsilon$ is excess noise. Hom: homodyne detection. Het: heterodyne detection. The side information transmitted from the classical authentication channel includes the selection results of Alice in postselection process and the information used in postprocessing process.}
	\label{fig:fig1-postselection}
\end{figure*}

Postselection algorithm can realize virtual physical operation, which greatly reduces the implementation complexity. However, the corresponding classical postprocessing part has not been deeply studied. There are still errors between Alice's and Bob's raw data after postselection. Therefore, it is necessary to correct the errors by using channel coding and decoding technology to obtain consistent keys. The raw data are continuous variables in CV-QKD. Therefore, it needs to be transformed into binary data that can be encoded through mapping transformation firstly. Generally, there are mainly two methods, slice reconciliation~\cite{lodewyck2007quantum} and multidimensional reconciliation~\cite{leverrier2008multidimensional}. The former is usually used in the case of high signal-to-noise ratio (SNR), while the latter is used in the case of low SNR. The common error correction codes used in CV-QKD are Raptor codes~\cite{shokrollahi2006raptor}, Polar codes~\cite{arikan2009channel}, and low density parity check codes (LDPC)~\cite{richardson2008modern} etc. Multi-edge type (MET) LDPC codes can achieve good error correction performance under extremely low SNRs.

In this paper, we mainly focus on the non-Gaussian reconciliation in CV-QKD. The raw data after photon subtraction no longer follow Gaussian distribution, the distribution varies with the number of photon subtraction. Firstly, we give the method which realizes the transformation from Gaussian distribution data to non Gaussian data through postselection filter function. This process is very important, which affects the amount of raw data saved after non-Gaussian postselection, it has an important impact on the secret key rate of CV-QKD system. Multidimensional reconciliation and MET-LDPC codes are used for the non-Gaussian reconciliation in reverse reconciliation system. We introduce layered belief propagation decoding algorithm~\cite{hocevar2004reduced} into non-Gaussian data error correction, which greatly reduces the complexity of the postprocessing decoding process and does not increase the frame error rate (FER) of decoding. Furthermore, we test the error correction performance of the non-Gaussian reconciliation. Although the noise and Bob's raw data still obey Gaussian distribution, Alice's data converge to medium amplitude after non-Gaussian postselection, so the anti-noise performance of the system is improved. It can be seen from the results that the FER of non-Gaussian data error correction is obviously lower than that of Gaussian data under the same conditions, the average number of iterations is also significantly reduced by using layered decoding algorithm.

The rest of the paper is organized as follows. In Sec. \ref{sec:2}, we introduce some basics of the non-Gaussian postselection, information reconciliation of CV-QKD, presenting a postselection method to convert Gaussian distribution data to non-Gaussian distribution data, proposing the non-Gaussian reconciliation algorithm based on multidimensional reconciliation and MET-LDPC codes. In Sec.~\ref{sec:3}, we present the data distribution under different virtual photon subtraction numbers, the performance tests of information reconciliation on the non-Gaussian data after postselection under different virtual photon subtraction numbers, and the error correction performance comparation with Gaussian data. In Sec.~\ref{sec:4}, we draw the conclusion of this paper.

\section{Non-Gaussian reconciliation in CV-QKD}
\label{sec:2}
Non-Gaussian operation can increase the entanglement of the Gaussian entangled states. As a non-Gaussian operation, it has been proposed that photon subtraction in CV-QKD can increase the transmission distance. It has been proved that the entanglement-based scheme and the corresponding prepare-and-measure scheme is secure~\cite{huang2013performance}. Simultaneously, the feasibility and security of virtual photon subtraction scheme through non-Gaussian postselection have also been proved~\cite{li2016non}. In this section, we first introduce the basic of the non-Gaussian postselection, then propose the postselection method to convert Gaussian distribution data to non-Gaussian distribution data, finally present the non-Gaussian reconciliation scheme based on multidimensional reconciliation and MET-LDPC codes.

\subsection{Non-Gaussian postselection in CV-QKD}
The security of entanglement-based model CV-QKD photon subtraction protocol has been proved, the security of the corresponding prepare-and-measure (PM) model with equivalent postselection as virtual photon subtraction has also been proved~\cite{li2016non}. Different from Gaussian quantum state protocol, non-Gaussian quantum state protocol can not have a symplectic covariance matrix, so we can’t use von-Neumann entropy to derive the Holevo boundary for secret key rate directly. However, we can estimate the secret key rate through the physical model equivalent to virtual photon subtraction. This equivalent physical model has been completed in our previous work~\cite{li2016non}. The main idea is based on the optimality of Gaussian attacks~\cite{garcia2006unconditional,41,42}. The eavesdropper will not get more information from the non-Gaussian quantum state than the Gaussian quantum state, so the secret key rate of the non-Gaussian protocol is less than that of the corresponding Gaussian protocol. To ensure unconditional security, we can use the secret key rate of the corresponding Gaussian protocol as the lower bound of the secret key rate of the non-Gaussian protocol. In addition, the probability of photon subtraction success should also be considered when estimating the secret key rate of non-Gaussian protocol.

Most of the implementation of CV-QKD is based on PM scheme, which does not need to prepare entangled states, so it is easy to implement in experiment. The PM scheme of CV-QKD with non-Gaussian postselection in Alice is shown in Fig.~\ref{fig:fig1-postselection}. 

In the PM scheme, Alice generates coherent states $\left| \alpha  \right\rangle$, where $\alpha = \sqrt{2T\lambda^2}\gamma $, $\gamma=x_A+ip_A$. $x_A$ and $p_A$ are both randomly selected from a Gaussian distribution data set with mean 0 and variance $V_A$. $T$ is related to the transmittance of photon subtraction, $\lambda^2=\dfrac{V-1}{V+1}$, and $V$ is variance of the two-mode squeezed vacuum state. Then after the coherent states are prepared, they are sent to Bob through quantum channels. Bob performs homodyne or heterodyne detection according to the type of protocol after receiving the quantum states, the measurement results are recorded as $x_B$ and $p_B$. If Bob performs homodyne detection, he informs Alice the quadratures ($x$ or $p$) he measures. Alice keeps the same quadratures with Bob. While in the case of heterodyne detection, both quadratures of $x$ and $p$ are kept.

After the quantum measurement and base sifting step, Alice chooses to accept part of the raw data according to the non-Gaussian postselection filter function with probability $P(k|x_A,p_A)$, the probability function of subtracting $k$ photons is described by
\begin{equation}\label{function}
	\begin{split}
		P(k|x_A,p_A)=\frac{1}{k!}[\dfrac{(1-T)\lambda^2}{2}(x_A^2+p_A^2)]^k\times\\
		\exp[-\dfrac{(1-T)\lambda^2}{2}(x_A^2+p_A^2)].
	\end{split}
\end{equation}

\begin{figure}[t]
	\centering
	\includegraphics[width=\linewidth]{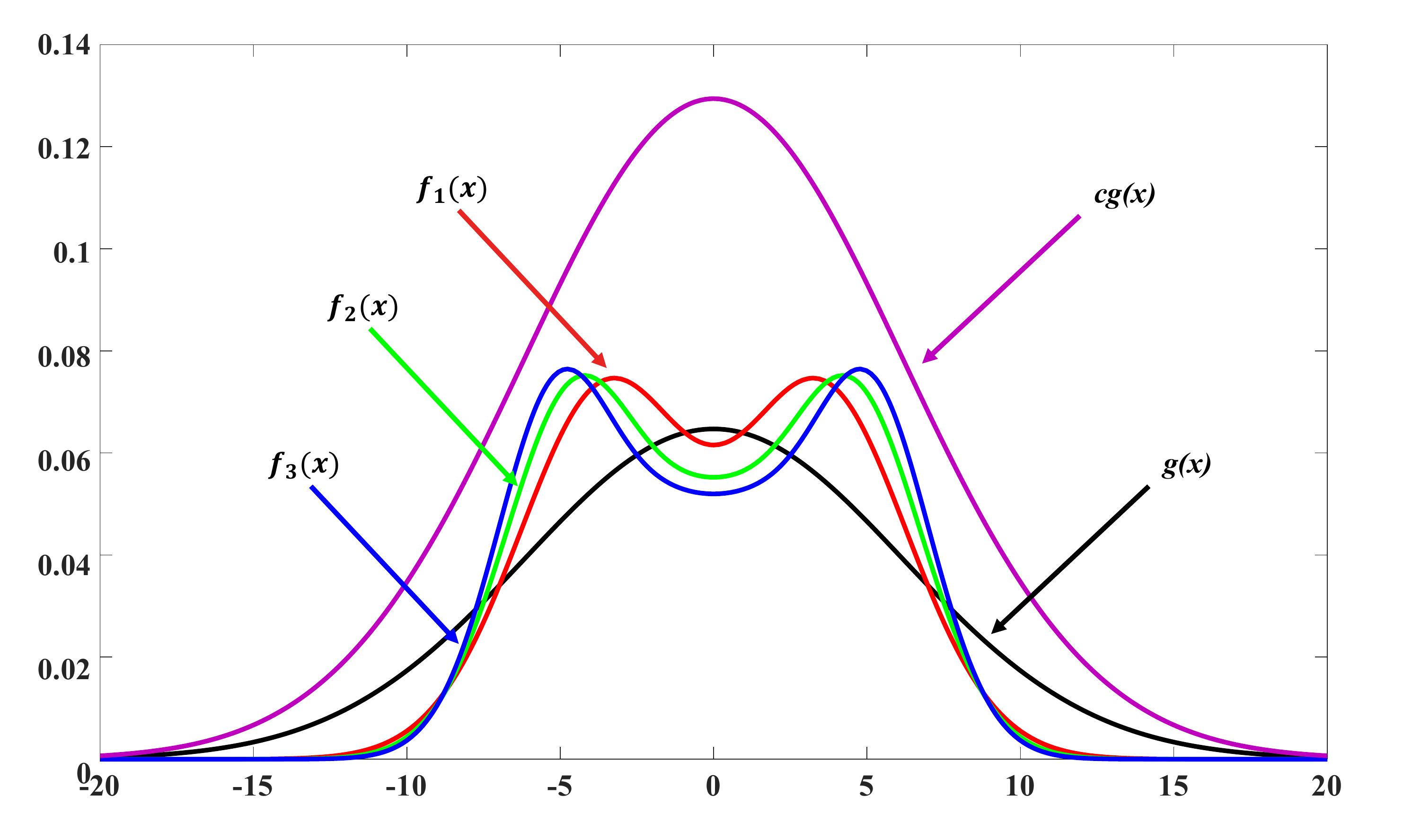}
	\caption{The probability density function of Gaussian distribution and Non-Gaussian distribution after postselection. The X-axis is the value of raw data, the Y-axis is the corresponding density. $g(x)$ is the probability density function of Gaussian distribution. $c$ is a constant for effective sampling. $f_1(x)$, $f_2(x)$, $f_3(x)$ are the probability density function of non-Gaussian distribution after virtual photon subtraction with $k=1, 2, 3$ separately. The modulation variance is set to 20.}
	\label{fig:fig2-distribution}
\end{figure}

Then she reveals the selection results to Bob. Bob keeps the corresponding raw data. After the non-Gaussian postselection, Alice's raw data no longer follow Gaussian distribution. But the distribution of Bob's raw data remains unchanged. The probability density function of the raw data before and after non-Gaussian postselection is shown in Fig.~\ref{fig:fig2-distribution}. The black solid line represents original Gaussian distribution function of Alice's raw data. The red, green, blue solid lines represent non-Gaussian distribution function of Alice's data after virtual photon substraction with $k=1, 2, 3$ respectively. As shown in Fig.~\ref{fig:fig2-distribution}, for the raw data of Alice, the probability of some data in Gaussian distribution is lower than that in non-Gaussian distribution. Therefore, it is impossible to directly sample non-Gaussian distribution data from Gaussian distribution data in this situation.

To solve this problem, the acceptance-rejection sampling method is used to convert the Gaussian distribution data to non-Gaussian distribution data, which is a sampling method from probability density function. This method is to extract subsequence of random variables from a sequence with a specific distribution according to a rule and make them meet the given probability distribution. Suppose the random variable is $x$, its value range is $[a,b]$, its probability density function $f(x)$ of non-Gaussian distribution is bounded, which satisfies max$\{f(x)|a\leqslant x\leqslant b\}=A.$ For effective sampling, the probability density function of Gaussian distribution is required to be greater than that of non-Gaussian distribution for any $x$. However, as shown in Fig.~\ref{fig:fig2-distribution}, the probability density function of raw data $g(x)$ does not satisfy the condition that it is always greater than $f(x)$ ($f_1(x)$, $f_2(x), f_3(x)$) for any $x$. In order to satisfy the condition, we construct a new probability density function which satisfies
\begin{equation}\label{cg}
	cg(x)\geqslant f_k(x), x\in[a,b],
\end{equation}
where $c$ is a constant. The purple solid line represents the probability density function of $cg(x)$, it can be seen that the probability is greater than that of $f(x)$ for any $x$. The value of $c$ should be as small as possible to improve sampling efficiency which can be defined as 
\begin{equation}\label{E}
	E=\dfrac{1}{m}.
\end{equation}

It represents the average number of original distribution random variables $m$ required to obtain a random variable with a specific distribution. In our case, it is equivalent to the success probability of the overall non-Gaussian postselection.

For a raw data $x$ of Alice, she randomly generates a random number $d$ that obeys a uniform distribution in the interval of $[0,cg(x)]$, where $d=cg(x)\xi, \xi \in U[0,1].$ Then, she compares $d$ and $f_k(x),$ if $d\leqslant f_k(x)$, she accepts $x$ as a non-Gaussian distribution data after virtual subtraction of $k$ photons. Otherwise, she rejects it and restarts the above process until all the raw data are completed.  

After completing the non-Gaussian postselection process, they use the saved data to perform postprocessing process through classical authentication channel, including information reconciliation, parameter estimation, and privacy amplification. The secret key rate against collective attacks for reverse reconciliation of the $k$ photons subtraction is described by 
\begin{equation}\label{SKR}
	K_{PS}^k=P(k)[\beta I(A:B)-S(E:B)],
\end{equation}
where $P(k)$ is the success probability of overall virtual $k$ photons subtraction (the success probability of the non-Gaussian postselection), $\beta$ is reconciliation efficiency, $I(A:B)$ is classical mutual information between Alice and Bob, $S(E:B)$ is Von Neumann entropy between Eve and Bob.

\subsection{Non-Gaussian reconciliation in CV-QKD}
Suppose that the variables of Alice and Bob are $X$ and $Y$ after non-Gaussian postselection. $X$ follows non-Gaussian distribution and $Y$ still follows Gaussian distribution. The quantum channel can be seen as an additive white Gaussian noise (AWGN) channel. Then we have $Y=tX+Z$, where $t$ is related to the channel transmittance and detection efficiency, $Z$ is channel noise which follows Gaussian distribution. For information reconciliation, signal to noise ratio (SNR) is the main parameter of concern. Thus for simplicity, we can fit $t=1$. In addition, direct reconciliation is limited by 3dB loss. Therefore, reverse reconciliation is used in our scheme, which can break this limit. Alice performs the error correction to obtain the identical keys with Bob.

Alice and Bob first convert the AWGN channel to a virtual binary input AWGN channel. Then they can use channel coding and decoding technology to correct errors between them. Multidimensional reconciliation is used to finish the first step. Bob normalizes the data $Y$ after postselection according to the dimension of multidimensional reconciliation. Then he randomly chooses a uniform distribution vector $U$ which is generated by a quantum random number generator. Next, a mapping function $M(Y',U)$ from normalized variable $Y'$ to $U$ is constructed by orthogonal transformation matrix. $Y'$ and $U$ satisfy the following relationship
\begin{equation}\label{MY}
	M(Y',U)Y'=U,
\end{equation}
He sends the mapping function to Alice through a classical authentication channel which means that eavesdropper Eve can get all the information but she can not change the information without the knowledge of both sides of the legal communication. Alice receives the mapping function $M(Y',U)$ and normalizes her non-Gaussian distribution data $X$ after postselection. Then she rotates her normalized data $X'$ to $V$ through the mapping function, which can be calculated by
\begin{equation}\label{MX}
	M(Y',U)X'=V,
\end{equation}
where $V$ is the noise form of $U$. They repeat the above process until all the data are converted. Finally, they select the appropriate channel codes according to the channel parameters to perform the error correction process.

MET-LDPC codes are the generalization form of LDPC codes, which is very suitable for error correction under extremely low SNRs. They can achieve performance close to the Shannon's limit. Thus, we choose MET-LDPC code as the channel coding technology to correct the errors between Alice and Bob. Firstly, The code rate of MET-LDPC code is calculated according to the estimated quantum channel characteristics. Secondly, the degree distribution of the code rate is obtained by density evolutionary algorithm. Then select a suitable construction method to generate the parity check matrix. Finally, Alice and Bob use the matrix for encoding and decoding to correct the errors between them to get completely consistent data.

The encoding process in CV-QKD system is very different from that in classical communication. It does not need generation matrix. The parity check matrix of error correction code is directly multiplied by the binary data after reconciliation to obtain the syndromes. Then Bob sends the syndromes to Alice through a classical authentication channel which will be errorless in the transmission process but Eve can get all the information. This is different from classical communication. After getting the syndromes, Alice initializes the information according to the data after reconciliation and uses the same parity check matrix to correct errors. In the message iteration process of error correction, the syndromes sent by Bob needs to be used and compared with Alice's temporary syndromes to judge whether the decoding is successful. This is another big difference from classical communication. 

The error correction codes used in CV-QKD are MET-LDPC codes, which have many decoding algorithms, such as probability domain belief propagation algorithm (BP), log likelihood ratio belief propagation algorithm (LLR-BP) and so on, in which BP algorithm is a commonly used decoding algorithm in CV-QKD system. Compared with other decoding algorithms, it has higher accuracy and can ensure the success rate of decoding in CV-QKD system. In the conventional BP decoding algorithm, after information initialization, it is necessary to traverse the check nodes and variable nodes to update the edge information in the parity check matrix, then updating the information of each node accordingly, finally comparing the syndromes. If the decision is successful, end the decoding, otherwise continue the iteration until the maximum number of iterations is reached. Although this algorithm has high accuracy, the updated nodes are in a waiting state before the next update, the utilization rate is low. Thus, a layered BP (LBP) decoding algorithm~\cite{hocevar2004reduced} is introduced to the postprocessing of CV-QKD system, which can make faster use of the updated node information and improve the decoding efficiency. In LBP decoding algorithm, each layer will update the variable node, taking the updated variable node as the input of the next layer to participate in the operation of the next layer in the same iteration. In this way, the updated information will be immediately used in this iteration, which can reduce the iteration number. Generally, it only needs half of the iteration number to achieve the same effect as the BP decoding algorithm. Thus, it can speed up the decoding process and it is suitable for application in high-speed postprocessing for CV-QKD system.

\section{Performance of the protocol}
\label{sec:3}
In this section, we first present the performance of the non-Gaussian postselection in terms of sampling efficiency from Gaussian distribution data to Non-Gaussian distribution data. Then we present the performance of non-Gaussian reconciliation for CV-QKD system in terms of reconciliation efficiency, frame error rate of error correction, and average iteration number. 

\subsection{Non-Gaussian postselection performance}
As described in Sec.~\ref{sec:2}, for the virtual $k$-photon subtraction, we cannot directly sample non-Gaussian data from the raw data of Gaussian distribution due to the probability of Gaussian distribution may be higher than that of non-Gaussian distribution. We present the acceptance-rejection sampling method to solve this problem, in which the sampling efficiency is a very important parameter to evaluate sampling performance. In addition to Eq.~\ref{E}, the sampling efficiency can also be expressed by geometric interpretation. It refers to the probability that the data conforming to $c_kg(x)$ falls under $f_k(x)$.

The selection of $c$ in Eq.~\ref{cg} has an important impact on sampling efficiency. First, it needs to satisfy that $f_k(x)$ is completely below $cg(x)$ in order to sample correctly. Secondly, the sampling efficiency should be as high as possible. That is, the value of $c_k$ should be as small as possible when condition 1 is met. Therefore, its value can be calculated by
\begin{equation}\label{ck}
	c_k=\max_{x}[\dfrac{f_k(x)}{g(x)}], x\in R.
\end{equation}

\begin{figure}[t]
	\centering
	\includegraphics[width=\linewidth]{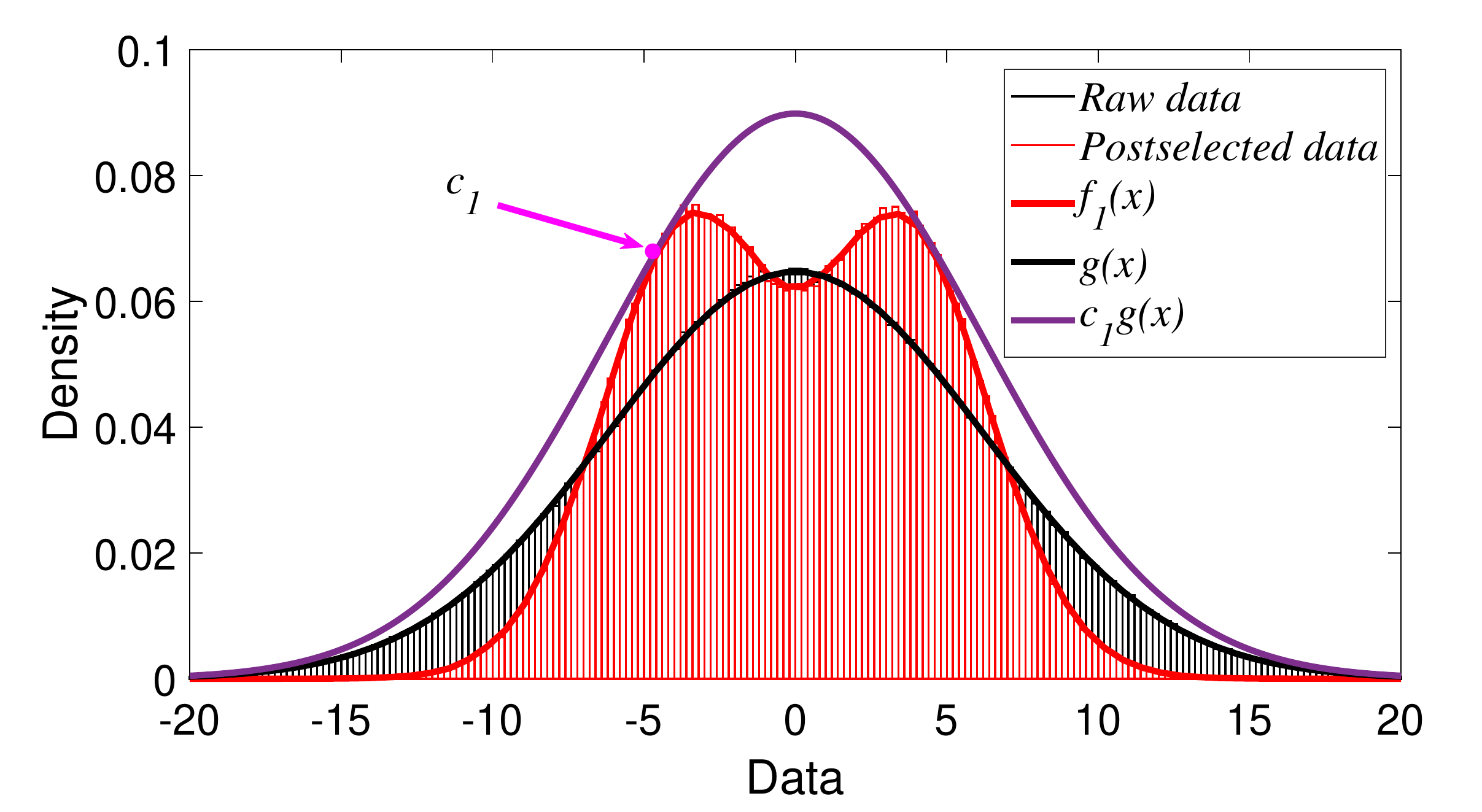}
	\caption{The Gaussian distribution probability
		density function of actual data and non-Gaussian probability density function with virtual subtraction of one photon. The thin black line represents the probability  density of raw data actually generated. The thin red line represents the probability  density of postselected data after virtual subtraction of one photon. $g(x)$ and  $f_1(x)$ are fitted probability density functions of raw data and postselected data (virtual subtraction of one photon) separately. The pink dot represents the point where $c$ gets the optimal value when virtual subtracting one photon, recording as $c_1$. The purple line represents the probability
		density function after $g(x)$ is enlarged by $c_1$ times. The modulation variance is set to 20.}
	\label{fig:fig3-ps1}
\end{figure}

As shown in Fig.~\ref{fig:fig3-ps1}, we give the Gaussian distribution probability density function of actual data and the corresponding non-Gaussian probability density function with virtual subtraction of one photon. We can obtain that the optimal value of $c_1$ is about 1.32 calculated by Eq.~\ref{ck} when $T$ is set to 0.8.

\noindent The corresponding sampling efficiency is about 75.4\%, which is much higher than the previous results of sampling with uniformly distributed data. Similarly, we can get the results of virtual $k$-photon subtraction.

\subsection{Non-Gaussian reconciliation performance}
Information reconciliation has an important impact on the performance of CV-QKD system. Reconciliation efficiency not only affects that whether secret keys can be extracted, but also affects the transmission distance of CV-QKD system. On the other hand, the success rate of reconciliation and processing speed also have an important impact on the secret key rate of the system. Thus, we test the error correction performance of both Gaussian and non-Gaussian data, including reconciliation efficiency, FER and average iteration number (AIN).

\begin{figure}[t]
	\centering
	\includegraphics[width=\linewidth]{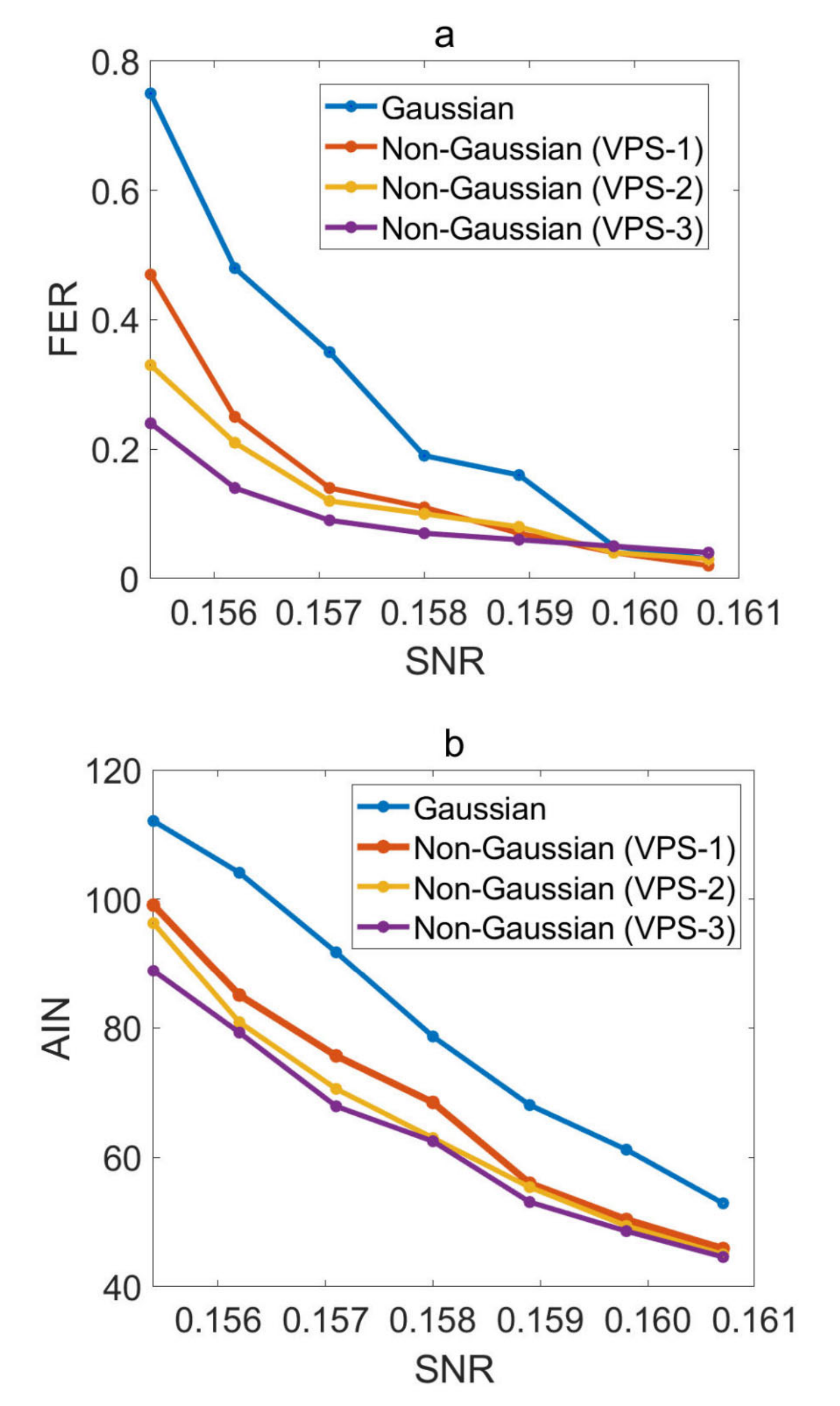}
	\caption{Performance comparison of error correction between Gaussian and non-Gaussian data. The error correction code used for Gaussian and non-Gaussian case is MET-LDPC code, whose rate is 0.1. (a) FER of Gaussian and non-Gaussian data after error correction under different SNR. (b) Average iteration number (AIN) of decoding corresponding to Gaussian and non-Gaussian data in (a). The blue line represents the FER/AIN of Gaussian data after error correction. The rest lines represent the FER/AIN of non-Gaussian data, where red line/yellow line/purple line respectively represents virtual 1-photon/2-photon/3-photon subtraction (VPS-1/VPS21/VPS-3). The dots represent the error correction performance of actual data, reconciliation efficiency from right to left is 93\%, 93.5\%, 94\%, 94.5\%, 95\%, 95.5\%, 96\% respectively. The maximum iteration number is set to 150.}
	\label{fig:fig4-FERAIN01}
\end{figure}

\begin{figure}[t]
	\centering
	\includegraphics[width=\linewidth]{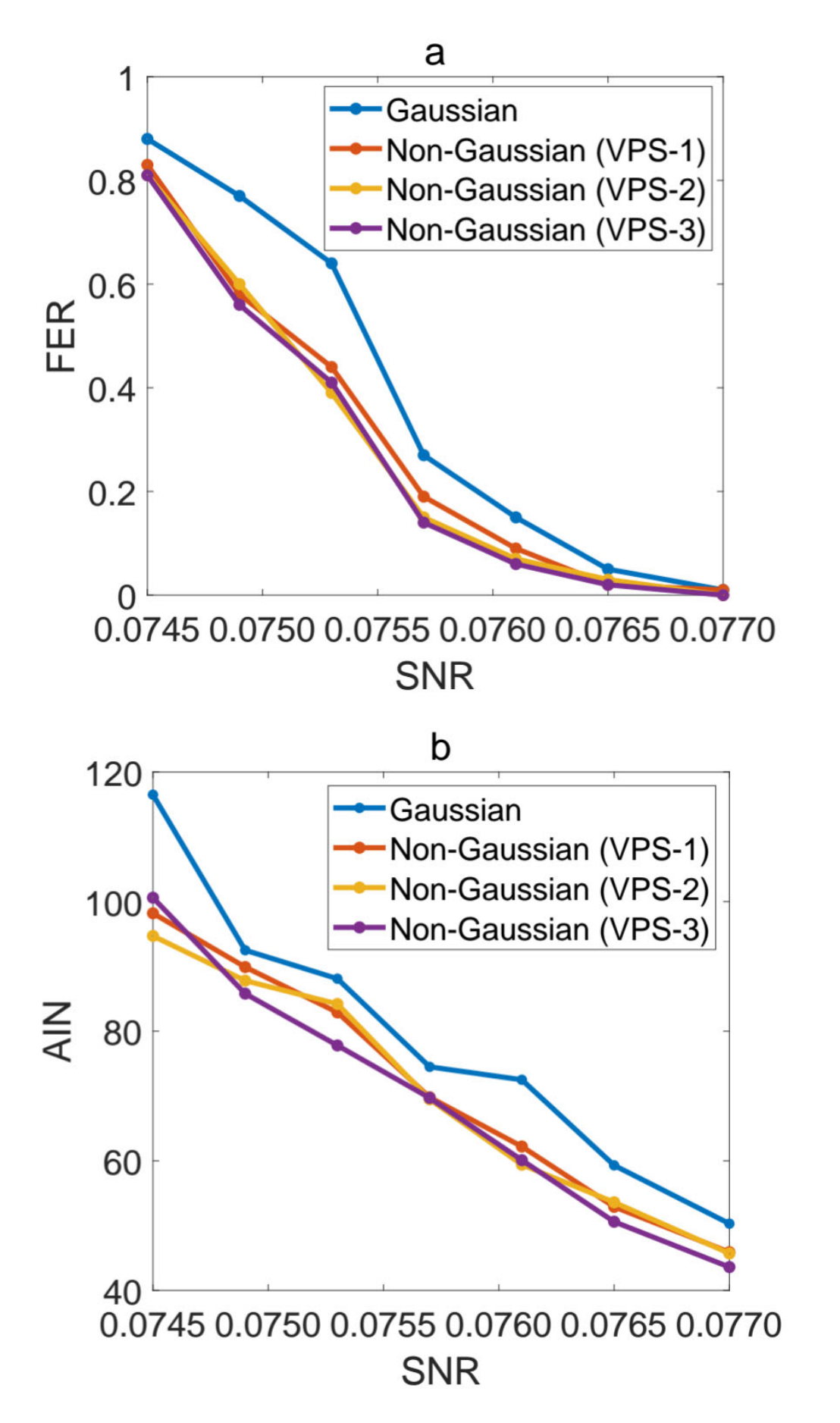}
	\caption{Performance comparison of error correction between Gaussian and non-Gaussian data under the code rate of 0.05. (a) FER of Gaussian and non-Gaussian data after error correction under different SNR. (b) Average iteration number (AIN) of decoding corresponding to Gaussian and non-Gaussian data in (a). The dots represent the error correction performance of actual data, reconciliation efficiency from right to left is 93.5\%, 94\%, 94.5\%, 95\%, 95.5\%, 96\%, 96.5\% respectively. The maximum iteration number is set to 150.}
	\label{fig:fig5-FERAIN005}
\end{figure}

The error correction performance of four types data of distribution is tested under two MET-LDPC codes. The data of these four types of distribution contains Gaussian distribution data and three non-Gaussian distribution data which includes the data of virtual 1-photon subtraction, virtual 2-photon subtraction, virtual 3-photon subtraction. The two MET-LDPC codes are rate of 0.1 and 0.05. For each rate of MET-LDPC code and each type of data, seven sets of data with different SNR are tested. The test results are shown in Fig.~\ref{fig:fig4-FERAIN01} and Fig.~\ref{fig:fig5-FERAIN005}.

We have tested more than 100 data block for each case. The size of each data block is $10^6$. As can be seen from the results of Fig.~\ref{fig:fig4-FERAIN01} and Fig.~\ref{fig:fig5-FERAIN005}, the error correction performance of non-Gaussian case is higher than that of Gaussian case both in FER and AIN. LBP decoding algorithm is used to the error correction process, which has little effect on FER of decoding. However it can reduce about half of the AIN, the decoding speed can be greatly increased to improve the secret key rate of CV-QKD system. FER is related to the set maximum number of iterations, the FER can be reduced by increasing the maximum number of iterations to a certain extent. But it will also increase the decoding delay simultaneously, thus there is a trade-off between FER and AIN. In order to compare the error correction performance under different conditions, we set the maximum number of iterations to 150 for all the tests. In actual applications, the maximum number of iterations can be reasonably set according to the average number of iterations for different cases.

Fig.~\ref{fig:fig4-FERAIN01} and Fig.~\ref{fig:fig5-FERAIN005} show the test results for two MET-LPDC codes, the code rate of 0.1 and 0.05 separately. Although the error correction performance of non-Gaussian data is better than that of Gaussian data at both codes,  it also have some different characteristics. For the code rate of 0.1, the error correction performance of non-Gaussian data is obviously better than that of Gaussian data both in FER and AIN, the error correction performance is better with the increase of the number of virtual photon subtraction, especially at a relatively low SNR. This is mainly due to the fact that non-Gaussian postselection diffuses the data originally concentrated around 0 to middle values, the larger the values diffusion with the increase of the number of virtual photon subtraction. Simultaneously, the large value will also concentrate to the middle value. Therefore, the error correction performance will not continue to increase. At a relatively high SNR, error correction is relatively easy at this time, so FER of error correction is very low (close to 0) for both Gaussian data and non-Gaussian data, continuing to increase the SNR cannot reflect the advantages of non-Gaussian data. However, non-Gaussian data error correction can converge faster, so the AIV is less than that of Gaussian data which will increase the error correction speed and the secret key rate of CV-QKD system. 

For the code rate of 0.05, the error correction performance of non-Gaussian data is also higher than that of Gaussian data, but its advantage is lower than that of 0.1 code rate matrix. This is because the SNR of error correction data corresponding to 0.05 code rate is low, the power of signal is much smaller than that of the noise. In this case, the signal is completely submerged in the noise, so it is difficult to correct the errors. Therefore, the advantages of non-Gaussian postselection are limited, with the increase of the number of photon subtraction, its advantage is not as obvious as that in 0.1 code rate. Although the error correction gain caused by different photon subtraction is not very obvious, the non-Gaussian postselection data still makes the decoding converge faster than that of Gaussian data due to the reduction of the values near 0. Therefore, the FER and AIN performance of decoding processing is still improved. 

We can use rate-adaptive method~\cite{wang2017efficient} or non-fixed rate error correction codes~\cite{zhou2019continuous} to expand the applicable range of SNR. We have studied these two methods in the preamble work, combining these two methods, the non-Gaussian error correction method proposed in this paper can play an advantage in a large range of SNR.

\begin{figure}[t]
	\centering
	\includegraphics[width=\linewidth]{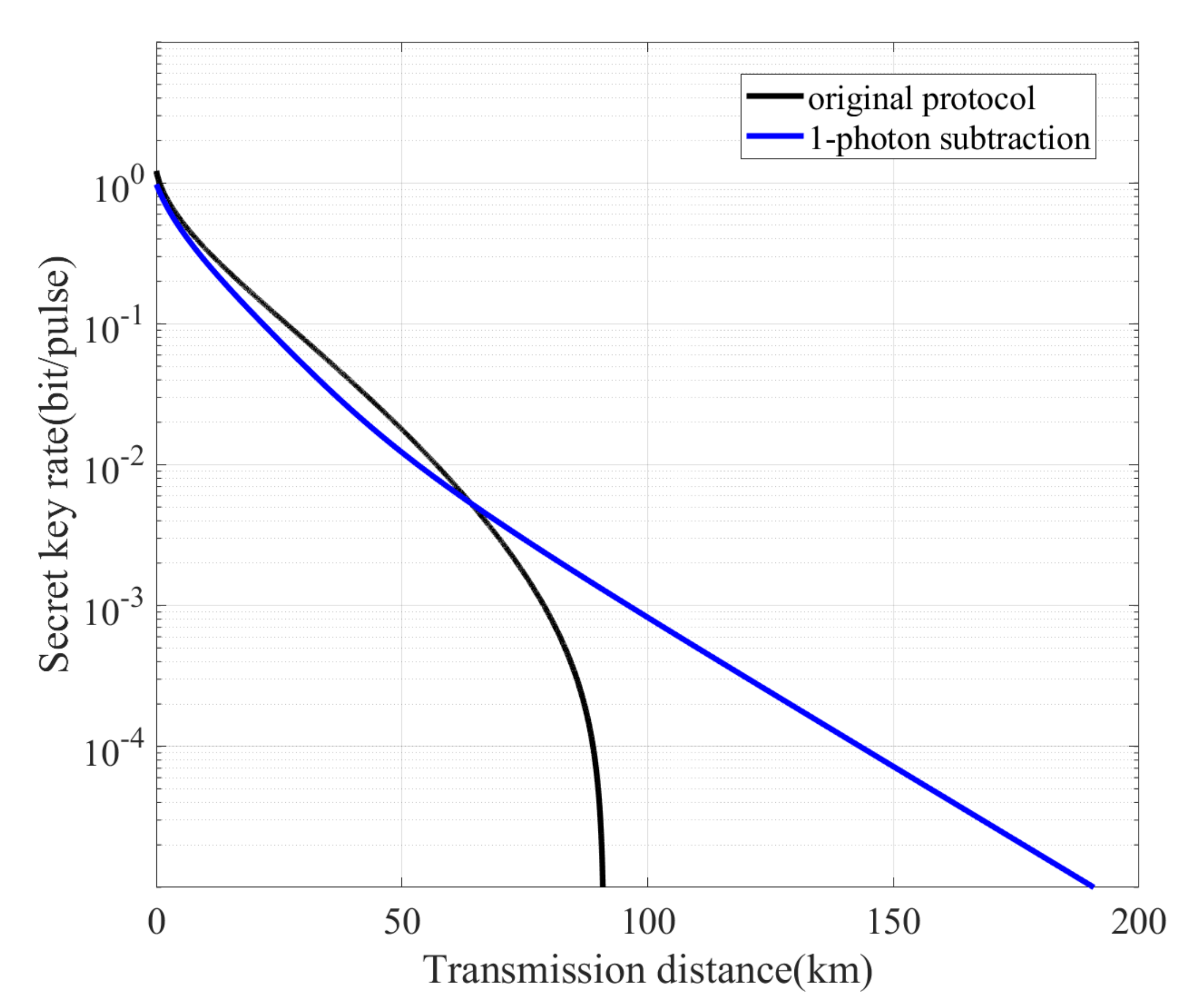}
	\caption{The secret key rate and transmission distance comparison of the original protocols (Gaussian case) and 1-photon subtraction (non-Gaussian case). The modulation variance is 20. Reconciliation efficiency is 95\%.}
	\label{fig:fig6-compare}
\end{figure}

Fig.~\ref{fig:fig6-compare} shows the secret key rate and transmission distance comparison of the original protocol (Gaussian case) and 1-photon subtraction protocol (non-Gaussian case). Fig.~\ref{fig:fig6-compare} shows that non-Gaussian protocols can still achieve high secret key rate over long distance range, which greatly expanding the maximal transmission distance. However, for short distance range, the secret key rate is worth than the original protocols. The main reason is that the probability of photon subtraction success is low. In other words, for non-Gaussian postselection, after selecting the original Gaussian data, since the amount of selected data is reduced, the average secret key rate is reduced.

In practical application, we are more concerned about the amount of secret keys obtained per unit time when the secret key rate is greater than 0. For non-Gaussian postselection, the reduction of data will lead to the reduction of the secret key rate of single pulse, but it will greatly improve the data processing speed. The reconciliation efficiency and decoding success rate of non-Gaussian data are higher than that of Gaussian data, thus the secret key rate per unit time is not necessarily lower than that in Gaussian case even at short distance range. It can be further studied in the subsequent high-speed implementation.

\section{Conclusion}
\label{sec:4}

In this paper, we proposed a non-Gaussian reconciliation method for CV-QKD protocols by non-Gaussian postselection at Alice's side, which can reduce the FER and AIN of decoding. We propose an effective postselection method to obtain non-Gaussian data follows specific distribution from Gaussian data, which greatly improves the success rate of virtual photon subtraction of CV-QKD system. Multidimensional reconciliation and MET-LDPC codes are used to perform the information reconciliation in postprocessing of CV-QKD system. The layered belief propagation decoding algorithm is introduced to the error correction, which can greatly reduce the decoding complexity and improve the decoding speed. We test the error correction performance of Gaussian data and non-Gaussian data after our proposed postselection under two representative codes with rate of 0.1 and 0.05. We test 7 sets of data for each case respectively. The corresponding reconciliation efficiency ranges from 93\% to 96.5\%. The results show that the FER and AIN of decoding performance of non-Gaussian data is significantly better than that of Gaussian data, which greatly improves the secret key rate of CV-QKD system. 

\begin{acknowledgments}
This work was supported by National Natural Science Foundation of China under Grant No. 62001041, No.62201012, the Fundamental Research Funds of BUPT under Grant No. 2022RC08, and the Fund of State Key Laboratory of Information Photonics and Optical Communications under Grant No. IPOC2022ZT09.

\end{acknowledgments}

\nocite{*}

\bibliography{sample}

\begin{thebibliography}{42}%
\makeatletter
\providecommand \@ifxundefined [1]{%
 \@ifx{#1\undefined}
}%
\providecommand \@ifnum [1]{%
 \ifnum #1\expandafter \@firstoftwo
 \else \expandafter \@secondoftwo
 \fi
}%
\providecommand \@ifx [1]{%
 \ifx #1\expandafter \@firstoftwo
 \else \expandafter \@secondoftwo
 \fi
}%
\providecommand \natexlab [1]{#1}%
\providecommand \enquote  [1]{``#1''}%
\providecommand \bibnamefont  [1]{#1}%
\providecommand \bibfnamefont [1]{#1}%
\providecommand \citenamefont [1]{#1}%
\providecommand \href@noop [0]{\@secondoftwo}%
\providecommand \href [0]{\begingroup \@sanitize@url \@href}%
\providecommand \@href[1]{\@@startlink{#1}\@@href}%
\providecommand \@@href[1]{\endgroup#1\@@endlink}%
\providecommand \@sanitize@url [0]{\catcode `\\12\catcode `\$12\catcode
  `\&12\catcode `\#12\catcode `\^12\catcode `\_12\catcode `\%12\relax}%
\providecommand \@@startlink[1]{}%
\providecommand \@@endlink[0]{}%
\providecommand \url  [0]{\begingroup\@sanitize@url \@url }%
\providecommand \@url [1]{\endgroup\@href {#1}{\urlprefix }}%
\providecommand \urlprefix  [0]{URL }%
\providecommand \Eprint [0]{\href }%
\providecommand \doibase [0]{https://doi.org/}%
\providecommand \selectlanguage [0]{\@gobble}%
\providecommand \bibinfo  [0]{\@secondoftwo}%
\providecommand \bibfield  [0]{\@secondoftwo}%
\providecommand \translation [1]{[#1]}%
\providecommand \BibitemOpen [0]{}%
\providecommand \bibitemStop [0]{}%
\providecommand \bibitemNoStop [0]{.\EOS\space}%
\providecommand \EOS [0]{\spacefactor3000\relax}%
\providecommand \BibitemShut  [1]{\csname bibitem#1\endcsname}%
\let\auto@bib@innerbib\@empty
\bibitem [{\citenamefont {Gisin}\ \emph {et~al.}(2002)\citenamefont {Gisin},
  \citenamefont {Ribordy}, \citenamefont {Tittel},\ and\ \citenamefont
  {Zbinden}}]{gisin2002quantum}%
  \BibitemOpen
  \bibfield  {author} {\bibinfo {author} {\bibfnamefont {N.}~\bibnamefont
  {Gisin}}, \bibinfo {author} {\bibfnamefont {G.}~\bibnamefont {Ribordy}},
  \bibinfo {author} {\bibfnamefont {W.}~\bibnamefont {Tittel}},\ and\ \bibinfo
  {author} {\bibfnamefont {H.}~\bibnamefont {Zbinden}},\ }\bibfield  {title}
  {\bibinfo {title} {Quantum cryptography},\ }\href
  {https://doi.org/10.1103/RevModPhys.74.145} {\bibfield  {journal} {\bibinfo
  {journal} {Rev. Mod. Phys.}\ }\textbf {\bibinfo {volume} {74}},\ \bibinfo
  {pages} {145} (\bibinfo {year} {2002})}\BibitemShut {NoStop}%
\bibitem [{\citenamefont {Xu}\ \emph {et~al.}(2020)\citenamefont {Xu},
  \citenamefont {Ma}, \citenamefont {Zhang}, \citenamefont {Lo},\ and\
  \citenamefont {Pan}}]{xu2020secure}%
  \BibitemOpen
  \bibfield  {author} {\bibinfo {author} {\bibfnamefont {F.}~\bibnamefont
  {Xu}}, \bibinfo {author} {\bibfnamefont {X.}~\bibnamefont {Ma}}, \bibinfo
  {author} {\bibfnamefont {Q.}~\bibnamefont {Zhang}}, \bibinfo {author}
  {\bibfnamefont {H.-K.}\ \bibnamefont {Lo}},\ and\ \bibinfo {author}
  {\bibfnamefont {J.-W.}\ \bibnamefont {Pan}},\ }\bibfield  {title} {\bibinfo
  {title} {Secure quantum key distribution with realistic devices},\ }\href
  {https://doi.org/10.1103/RevModPhys.92.025002} {\bibfield  {journal}
  {\bibinfo  {journal} {Rev. Mod. Phys.}\ }\textbf {\bibinfo {volume} {92}},\
  \bibinfo {pages} {025002} (\bibinfo {year} {2020})}\BibitemShut {NoStop}%
\bibitem [{\citenamefont {Pirandola}\ \emph {et~al.}(2020)\citenamefont
  {Pirandola}, \citenamefont {Andersen}, \citenamefont {Banchi}, \citenamefont
  {Berta}, \citenamefont {Bunandar}, \citenamefont {Colbeck}, \citenamefont
  {Englund}, \citenamefont {Gehring}, \citenamefont {Lupo}, \citenamefont
  {Ottaviani}, \citenamefont {Pereira}, \citenamefont {Razavi}, \citenamefont
  {Shaari}, \citenamefont {Tomamichel}, \citenamefont {Usenko}, \citenamefont
  {Vallone}, \citenamefont {Villoresi},\ and\ \citenamefont
  {Wallden}}]{pirandola2020advances}%
  \BibitemOpen
  \bibfield  {author} {\bibinfo {author} {\bibfnamefont {S.}~\bibnamefont
  {Pirandola}}, \bibinfo {author} {\bibfnamefont {U.~L.}\ \bibnamefont
  {Andersen}}, \bibinfo {author} {\bibfnamefont {L.}~\bibnamefont {Banchi}},
  \bibinfo {author} {\bibfnamefont {M.}~\bibnamefont {Berta}}, \bibinfo
  {author} {\bibfnamefont {D.}~\bibnamefont {Bunandar}}, \bibinfo {author}
  {\bibfnamefont {R.}~\bibnamefont {Colbeck}}, \bibinfo {author} {\bibfnamefont
  {D.}~\bibnamefont {Englund}}, \bibinfo {author} {\bibfnamefont
  {T.}~\bibnamefont {Gehring}}, \bibinfo {author} {\bibfnamefont
  {C.}~\bibnamefont {Lupo}}, \bibinfo {author} {\bibfnamefont {C.}~\bibnamefont
  {Ottaviani}}, \bibinfo {author} {\bibfnamefont {J.~L.}\ \bibnamefont
  {Pereira}}, \bibinfo {author} {\bibfnamefont {M.}~\bibnamefont {Razavi}},
  \bibinfo {author} {\bibfnamefont {J.~S.}\ \bibnamefont {Shaari}}, \bibinfo
  {author} {\bibfnamefont {M.}~\bibnamefont {Tomamichel}}, \bibinfo {author}
  {\bibfnamefont {V.~C.}\ \bibnamefont {Usenko}}, \bibinfo {author}
  {\bibfnamefont {G.}~\bibnamefont {Vallone}}, \bibinfo {author} {\bibfnamefont
  {P.}~\bibnamefont {Villoresi}},\ and\ \bibinfo {author} {\bibfnamefont
  {P.}~\bibnamefont {Wallden}},\ }\bibfield  {title} {\bibinfo {title}
  {Advances in quantum cryptography},\ }\href
  {https://doi.org/10.1364/AOP.361502} {\bibfield  {journal} {\bibinfo
  {journal} {Adv. Opt. Photon.}\ }\textbf {\bibinfo {volume} {12}},\ \bibinfo
  {pages} {1012} (\bibinfo {year} {2020})}\BibitemShut {NoStop}%
\bibitem [{\citenamefont {Bennett}(1984)}]{bennett1984quantum}%
  \BibitemOpen
  \bibfield  {author} {\bibinfo {author} {\bibfnamefont {C.~H.}\ \bibnamefont
  {Bennett}},\ }\bibfield  {title} {\bibinfo {title} {Quantum cryptography:
  Public key distribution and coin tossing},\ }in\ \href@noop {} {\emph
  {\bibinfo {booktitle} {Proc of IEEE International Conference on Computers}}}\
  (\bibinfo {year} {1984})\BibitemShut {NoStop}%
\bibitem [{\citenamefont {Wang}\ \emph
  {et~al.}(2022{\natexlab{a}})\citenamefont {Wang}, \citenamefont {Yin},
  \citenamefont {He}, \citenamefont {Chen}, \citenamefont {Wang}, \citenamefont
  {Ye}, \citenamefont {Zhou}, \citenamefont {Fan-Yuan}, \citenamefont {Wang},
  \citenamefont {Chen} \emph {et~al.}}]{wang2022twin}%
  \BibitemOpen
  \bibfield  {author} {\bibinfo {author} {\bibfnamefont {S.}~\bibnamefont
  {Wang}}, \bibinfo {author} {\bibfnamefont {Z.-Q.}\ \bibnamefont {Yin}},
  \bibinfo {author} {\bibfnamefont {D.-Y.}\ \bibnamefont {He}}, \bibinfo
  {author} {\bibfnamefont {W.}~\bibnamefont {Chen}}, \bibinfo {author}
  {\bibfnamefont {R.-Q.}\ \bibnamefont {Wang}}, \bibinfo {author}
  {\bibfnamefont {P.}~\bibnamefont {Ye}}, \bibinfo {author} {\bibfnamefont
  {Y.}~\bibnamefont {Zhou}}, \bibinfo {author} {\bibfnamefont {G.-J.}\
  \bibnamefont {Fan-Yuan}}, \bibinfo {author} {\bibfnamefont {F.-X.}\
  \bibnamefont {Wang}}, \bibinfo {author} {\bibfnamefont {W.}~\bibnamefont
  {Chen}}, \emph {et~al.},\ }\bibfield  {title} {\bibinfo {title} {Twin-field
  quantum key distribution over 830-km fibre},\ }\href
  {https://www.nature.com/articles/s41566-021-00928-2} {\bibfield  {journal}
  {\bibinfo  {journal} {Nature Photonics}\ }\textbf {\bibinfo {volume} {16}},\
  \bibinfo {pages} {154} (\bibinfo {year} {2022}{\natexlab{a}})}\BibitemShut
  {NoStop}%
\bibitem [{\citenamefont {Grosshans}\ and\ \citenamefont
  {Grangier}(2002)}]{grosshans2002continuous}%
  \BibitemOpen
  \bibfield  {author} {\bibinfo {author} {\bibfnamefont {F.}~\bibnamefont
  {Grosshans}}\ and\ \bibinfo {author} {\bibfnamefont {P.}~\bibnamefont
  {Grangier}},\ }\bibfield  {title} {\bibinfo {title} {Continuous variable
  quantum cryptography using coherent states},\ }\href
  {https://doi.org/10.1103/PhysRevLett.88.057902} {\bibfield  {journal}
  {\bibinfo  {journal} {Phys. Rev. Lett.}\ }\textbf {\bibinfo {volume} {88}},\
  \bibinfo {pages} {057902} (\bibinfo {year} {2002})}\BibitemShut {NoStop}%
\bibitem [{\citenamefont {Grosshans}\ \emph {et~al.}(2003)\citenamefont
  {Grosshans}, \citenamefont {Van~Assche}, \citenamefont {Wenger},
  \citenamefont {Brouri}, \citenamefont {Cerf},\ and\ \citenamefont
  {Grangier}}]{grosshans2003quantum}%
  \BibitemOpen
  \bibfield  {author} {\bibinfo {author} {\bibfnamefont {F.}~\bibnamefont
  {Grosshans}}, \bibinfo {author} {\bibfnamefont {G.}~\bibnamefont
  {Van~Assche}}, \bibinfo {author} {\bibfnamefont {J.}~\bibnamefont {Wenger}},
  \bibinfo {author} {\bibfnamefont {R.}~\bibnamefont {Brouri}}, \bibinfo
  {author} {\bibfnamefont {N.~J.}\ \bibnamefont {Cerf}},\ and\ \bibinfo
  {author} {\bibfnamefont {P.}~\bibnamefont {Grangier}},\ }\bibfield  {title}
  {\bibinfo {title} {Quantum key distribution using gaussian-modulated coherent
  states},\ }\href {https://www.nature.com/articles/nature01289} {\bibfield
  {journal} {\bibinfo  {journal} {Nature}\ }\textbf {\bibinfo {volume} {421}},\
  \bibinfo {pages} {238} (\bibinfo {year} {2003})}\BibitemShut {NoStop}%
\bibitem [{\citenamefont {Weedbrook}\ \emph {et~al.}(2012)\citenamefont
  {Weedbrook}, \citenamefont {Pirandola}, \citenamefont {Garc\'{\i}a-Patr\'on},
  \citenamefont {Cerf}, \citenamefont {Ralph}, \citenamefont {Shapiro},\ and\
  \citenamefont {Lloyd}}]{weedbrook2012gaussian}%
  \BibitemOpen
  \bibfield  {author} {\bibinfo {author} {\bibfnamefont {C.}~\bibnamefont
  {Weedbrook}}, \bibinfo {author} {\bibfnamefont {S.}~\bibnamefont
  {Pirandola}}, \bibinfo {author} {\bibfnamefont {R.}~\bibnamefont
  {Garc\'{\i}a-Patr\'on}}, \bibinfo {author} {\bibfnamefont {N.~J.}\
  \bibnamefont {Cerf}}, \bibinfo {author} {\bibfnamefont {T.~C.}\ \bibnamefont
  {Ralph}}, \bibinfo {author} {\bibfnamefont {J.~H.}\ \bibnamefont {Shapiro}},\
  and\ \bibinfo {author} {\bibfnamefont {S.}~\bibnamefont {Lloyd}},\ }\bibfield
   {title} {\bibinfo {title} {Gaussian quantum information},\ }\href
  {https://doi.org/10.1103/RevModPhys.84.621} {\bibfield  {journal} {\bibinfo
  {journal} {Rev. Mod. Phys.}\ }\textbf {\bibinfo {volume} {84}},\ \bibinfo
  {pages} {621} (\bibinfo {year} {2012})}\BibitemShut {NoStop}%
\bibitem [{\citenamefont {Jouguet}\ \emph {et~al.}(2013)\citenamefont
  {Jouguet}, \citenamefont {Kunz-Jacques}, \citenamefont {Leverrier},
  \citenamefont {Grangier},\ and\ \citenamefont
  {Diamanti}}]{jouguet2013experimental}%
  \BibitemOpen
  \bibfield  {author} {\bibinfo {author} {\bibfnamefont {P.}~\bibnamefont
  {Jouguet}}, \bibinfo {author} {\bibfnamefont {S.}~\bibnamefont
  {Kunz-Jacques}}, \bibinfo {author} {\bibfnamefont {A.}~\bibnamefont
  {Leverrier}}, \bibinfo {author} {\bibfnamefont {P.}~\bibnamefont
  {Grangier}},\ and\ \bibinfo {author} {\bibfnamefont {E.}~\bibnamefont
  {Diamanti}},\ }\bibfield  {title} {\bibinfo {title} {Experimental
  demonstration of long-distance continuous-variable quantum key
  distribution},\ }\href {https://www.nature.com/articles/nphoton.2013.63}
  {\bibfield  {journal} {\bibinfo  {journal} {Nature photonics}\ }\textbf
  {\bibinfo {volume} {7}},\ \bibinfo {pages} {378} (\bibinfo {year}
  {2013})}\BibitemShut {NoStop}%
\bibitem [{\citenamefont {Garc\'{\i}a-Patr\'on}\ and\ \citenamefont
  {Cerf}(2006)}]{garcia2006unconditional}%
  \BibitemOpen
  \bibfield  {author} {\bibinfo {author} {\bibfnamefont {R.}~\bibnamefont
  {Garc\'{\i}a-Patr\'on}}\ and\ \bibinfo {author} {\bibfnamefont {N.~J.}\
  \bibnamefont {Cerf}},\ }\bibfield  {title} {\bibinfo {title} {Unconditional
  optimality of gaussian attacks against continuous-variable quantum key
  distribution},\ }\href {https://doi.org/10.1103/PhysRevLett.97.190503}
  {\bibfield  {journal} {\bibinfo  {journal} {Phys. Rev. Lett.}\ }\textbf
  {\bibinfo {volume} {97}},\ \bibinfo {pages} {190503} (\bibinfo {year}
  {2006})}\BibitemShut {NoStop}%
\bibitem [{\citenamefont {Furrer}\ \emph {et~al.}(2012)\citenamefont {Furrer},
  \citenamefont {Franz}, \citenamefont {Berta}, \citenamefont {Leverrier},
  \citenamefont {Scholz}, \citenamefont {Tomamichel},\ and\ \citenamefont
  {Werner}}]{furrer2012continuous}%
  \BibitemOpen
  \bibfield  {author} {\bibinfo {author} {\bibfnamefont {F.}~\bibnamefont
  {Furrer}}, \bibinfo {author} {\bibfnamefont {T.}~\bibnamefont {Franz}},
  \bibinfo {author} {\bibfnamefont {M.}~\bibnamefont {Berta}}, \bibinfo
  {author} {\bibfnamefont {A.}~\bibnamefont {Leverrier}}, \bibinfo {author}
  {\bibfnamefont {V.~B.}\ \bibnamefont {Scholz}}, \bibinfo {author}
  {\bibfnamefont {M.}~\bibnamefont {Tomamichel}},\ and\ \bibinfo {author}
  {\bibfnamefont {R.~F.}\ \bibnamefont {Werner}},\ }\bibfield  {title}
  {\bibinfo {title} {Continuous variable quantum key distribution: Finite-key
  analysis of composable security against coherent attacks},\ }\href
  {https://doi.org/10.1103/PhysRevLett.109.100502} {\bibfield  {journal}
  {\bibinfo  {journal} {Phys. Rev. Lett.}\ }\textbf {\bibinfo {volume} {109}},\
  \bibinfo {pages} {100502} (\bibinfo {year} {2012})}\BibitemShut {NoStop}%
\bibitem [{\citenamefont {Leverrier}(2015)}]{leverrier2015composable}%
  \BibitemOpen
  \bibfield  {author} {\bibinfo {author} {\bibfnamefont {A.}~\bibnamefont
  {Leverrier}},\ }\bibfield  {title} {\bibinfo {title} {Composable security
  proof for continuous-variable quantum key distribution with coherent
  states},\ }\href {https://doi.org/10.1103/PhysRevLett.114.070501} {\bibfield
  {journal} {\bibinfo  {journal} {Phys. Rev. Lett.}\ }\textbf {\bibinfo
  {volume} {114}},\ \bibinfo {pages} {070501} (\bibinfo {year}
  {2015})}\BibitemShut {NoStop}%
\bibitem [{\citenamefont {Jain}\ \emph {et~al.}(2022)\citenamefont {Jain},
  \citenamefont {Chin}, \citenamefont {Mani}, \citenamefont {Lupo},
  \citenamefont {Nikolic}, \citenamefont {Kordts}, \citenamefont {Pirandola},
  \citenamefont {Pedersen}, \citenamefont {Kolb}, \citenamefont {{\"O}mer}
  \emph {et~al.}}]{jain2022practical}%
  \BibitemOpen
  \bibfield  {author} {\bibinfo {author} {\bibfnamefont {N.}~\bibnamefont
  {Jain}}, \bibinfo {author} {\bibfnamefont {H.-M.}\ \bibnamefont {Chin}},
  \bibinfo {author} {\bibfnamefont {H.}~\bibnamefont {Mani}}, \bibinfo {author}
  {\bibfnamefont {C.}~\bibnamefont {Lupo}}, \bibinfo {author} {\bibfnamefont
  {D.~S.}\ \bibnamefont {Nikolic}}, \bibinfo {author} {\bibfnamefont
  {A.}~\bibnamefont {Kordts}}, \bibinfo {author} {\bibfnamefont
  {S.}~\bibnamefont {Pirandola}}, \bibinfo {author} {\bibfnamefont {T.~B.}\
  \bibnamefont {Pedersen}}, \bibinfo {author} {\bibfnamefont {M.}~\bibnamefont
  {Kolb}}, \bibinfo {author} {\bibfnamefont {B.}~\bibnamefont {{\"O}mer}},
  \emph {et~al.},\ }\bibfield  {title} {\bibinfo {title} {Practical
  continuous-variable quantum key distribution with composable security},\
  }\href {https://doi.org/10.1038/s41467-022-32161-y} {\bibfield  {journal}
  {\bibinfo  {journal} {Nature communications}\ }\textbf {\bibinfo {volume}
  {13}},\ \bibinfo {pages} {4740} (\bibinfo {year} {2022})}\BibitemShut
  {NoStop}%
\bibitem [{\citenamefont {Chen}\ \emph {et~al.}(2023)\citenamefont {Chen},
  \citenamefont {Wang}, \citenamefont {Yu}, \citenamefont {Li},\ and\
  \citenamefont {Guo}}]{chen2023continuous}%
  \BibitemOpen
  \bibfield  {author} {\bibinfo {author} {\bibfnamefont {Z.}~\bibnamefont
  {Chen}}, \bibinfo {author} {\bibfnamefont {X.}~\bibnamefont {Wang}}, \bibinfo
  {author} {\bibfnamefont {S.}~\bibnamefont {Yu}}, \bibinfo {author}
  {\bibfnamefont {Z.}~\bibnamefont {Li}},\ and\ \bibinfo {author}
  {\bibfnamefont {H.}~\bibnamefont {Guo}},\ }\bibfield  {title} {\bibinfo
  {title} {Continuous-mode quantum key distribution with digital signal
  processing},\ }\href {https://doi.org/10.1038/s41534-023-00695-8} {\bibfield
  {journal} {\bibinfo  {journal} {npj Quantum Information}\ }\textbf {\bibinfo
  {volume} {9}},\ \bibinfo {pages} {28} (\bibinfo {year} {2023})}\BibitemShut
  {NoStop}%
\bibitem [{\citenamefont {Wang}\ \emph {et~al.}(2020)\citenamefont {Wang},
  \citenamefont {Pi}, \citenamefont {Huang}, \citenamefont {Li}, \citenamefont
  {Shao}, \citenamefont {Yang}, \citenamefont {Liu}, \citenamefont {Zhang},
  \citenamefont {Zhang},\ and\ \citenamefont {Xu}}]{wang2020high}%
  \BibitemOpen
  \bibfield  {author} {\bibinfo {author} {\bibfnamefont {H.}~\bibnamefont
  {Wang}}, \bibinfo {author} {\bibfnamefont {Y.}~\bibnamefont {Pi}}, \bibinfo
  {author} {\bibfnamefont {W.}~\bibnamefont {Huang}}, \bibinfo {author}
  {\bibfnamefont {Y.}~\bibnamefont {Li}}, \bibinfo {author} {\bibfnamefont
  {Y.}~\bibnamefont {Shao}}, \bibinfo {author} {\bibfnamefont {J.}~\bibnamefont
  {Yang}}, \bibinfo {author} {\bibfnamefont {J.}~\bibnamefont {Liu}}, \bibinfo
  {author} {\bibfnamefont {C.}~\bibnamefont {Zhang}}, \bibinfo {author}
  {\bibfnamefont {Y.}~\bibnamefont {Zhang}},\ and\ \bibinfo {author}
  {\bibfnamefont {B.}~\bibnamefont {Xu}},\ }\bibfield  {title} {\bibinfo
  {title} {High-speed gaussian-modulated continuous-variable quantum key
  distribution with a local local oscillator based on pilot-tone-assisted phase
  compensation},\ }\href {https://doi.org/10.1364/OE.404611} {\bibfield
  {journal} {\bibinfo  {journal} {Opt. Express}\ }\textbf {\bibinfo {volume}
  {28}},\ \bibinfo {pages} {32882} (\bibinfo {year} {2020})}\BibitemShut
  {NoStop}%
\bibitem [{\citenamefont {Shen}\ \emph {et~al.}(2021)\citenamefont {Shen},
  \citenamefont {Huang}, \citenamefont {Wang}, \citenamefont {Tian},
  \citenamefont {Chen},\ and\ \citenamefont {Yu}}]{shen2021strengthening}%
  \BibitemOpen
  \bibfield  {author} {\bibinfo {author} {\bibfnamefont {T.}~\bibnamefont
  {Shen}}, \bibinfo {author} {\bibfnamefont {Y.}~\bibnamefont {Huang}},
  \bibinfo {author} {\bibfnamefont {X.}~\bibnamefont {Wang}}, \bibinfo {author}
  {\bibfnamefont {H.}~\bibnamefont {Tian}}, \bibinfo {author} {\bibfnamefont
  {Z.}~\bibnamefont {Chen}},\ and\ \bibinfo {author} {\bibfnamefont
  {S.}~\bibnamefont {Yu}},\ }\bibfield  {title} {\bibinfo {title}
  {Strengthening practical continuous-variable quantum key distribution against
  measurement angular error},\ }\href {https://doi.org/10.1364/OE.433576}
  {\bibfield  {journal} {\bibinfo  {journal} {Opt. Express}\ }\textbf {\bibinfo
  {volume} {29}},\ \bibinfo {pages} {30978} (\bibinfo {year}
  {2021})}\BibitemShut {NoStop}%
\bibitem [{\citenamefont {Wang}\ \emph {et~al.}(2017)\citenamefont {Wang},
  \citenamefont {Zhang}, \citenamefont {Yu}, \citenamefont {Xu}, \citenamefont
  {Li},\ and\ \citenamefont {Guo}}]{wang2017efficient}%
  \BibitemOpen
  \bibfield  {author} {\bibinfo {author} {\bibfnamefont {X.}~\bibnamefont
  {Wang}}, \bibinfo {author} {\bibfnamefont {Y.}~\bibnamefont {Zhang}},
  \bibinfo {author} {\bibfnamefont {S.}~\bibnamefont {Yu}}, \bibinfo {author}
  {\bibfnamefont {B.}~\bibnamefont {Xu}}, \bibinfo {author} {\bibfnamefont
  {Z.}~\bibnamefont {Li}},\ and\ \bibinfo {author} {\bibfnamefont
  {H.}~\bibnamefont {Guo}},\ }\bibfield  {title} {\bibinfo {title} {Efficient
  rate-adaptive reconciliation for continuous-variable quantum key
  distribution},\ }\href {https://doi.org/10.26421/QIC17.13-14-4} {\bibfield
  {journal} {\bibinfo  {journal} {Quantum Inf. Comput.}\ }\textbf {\bibinfo
  {volume} {17}},\ \bibinfo {pages} {1123} (\bibinfo {year}
  {2017})}\BibitemShut {NoStop}%
\bibitem [{\citenamefont {Wang}\ \emph {et~al.}(2019)\citenamefont {Wang},
  \citenamefont {Guo}, \citenamefont {Wang}, \citenamefont {Liu},\ and\
  \citenamefont {Li}}]{wang2019realistic}%
  \BibitemOpen
  \bibfield  {author} {\bibinfo {author} {\bibfnamefont {X.}~\bibnamefont
  {Wang}}, \bibinfo {author} {\bibfnamefont {S.}~\bibnamefont {Guo}}, \bibinfo
  {author} {\bibfnamefont {P.}~\bibnamefont {Wang}}, \bibinfo {author}
  {\bibfnamefont {W.}~\bibnamefont {Liu}},\ and\ \bibinfo {author}
  {\bibfnamefont {Y.}~\bibnamefont {Li}},\ }\bibfield  {title} {\bibinfo
  {title} {Realistic rate-distance limit of continuous-variable quantum key
  distribution},\ }\href {https://doi.org/10.1364/OE.27.013372} {\bibfield
  {journal} {\bibinfo  {journal} {Opt. Express}\ }\textbf {\bibinfo {volume}
  {27}},\ \bibinfo {pages} {13372} (\bibinfo {year} {2019})}\BibitemShut
  {NoStop}%
\bibitem [{\citenamefont {Huang}\ \emph {et~al.}(2015)\citenamefont {Huang},
  \citenamefont {Lin}, \citenamefont {Wang}, \citenamefont {Liu}, \citenamefont
  {Fang}, \citenamefont {Peng}, \citenamefont {Huang},\ and\ \citenamefont
  {Zeng}}]{huang2015continuous}%
  \BibitemOpen
  \bibfield  {author} {\bibinfo {author} {\bibfnamefont {D.}~\bibnamefont
  {Huang}}, \bibinfo {author} {\bibfnamefont {D.}~\bibnamefont {Lin}}, \bibinfo
  {author} {\bibfnamefont {C.}~\bibnamefont {Wang}}, \bibinfo {author}
  {\bibfnamefont {W.}~\bibnamefont {Liu}}, \bibinfo {author} {\bibfnamefont
  {S.}~\bibnamefont {Fang}}, \bibinfo {author} {\bibfnamefont {J.}~\bibnamefont
  {Peng}}, \bibinfo {author} {\bibfnamefont {P.}~\bibnamefont {Huang}},\ and\
  \bibinfo {author} {\bibfnamefont {G.}~\bibnamefont {Zeng}},\ }\bibfield
  {title} {\bibinfo {title} {Continuous-variable quantum key distribution with
  1 mbps secure key rate},\ }\href {https://doi.org/10.1364/OE.23.017511}
  {\bibfield  {journal} {\bibinfo  {journal} {Opt. Express}\ }\textbf {\bibinfo
  {volume} {23}},\ \bibinfo {pages} {17511} (\bibinfo {year}
  {2015})}\BibitemShut {NoStop}%
\bibitem [{\citenamefont {Wang}\ \emph
  {et~al.}(2018{\natexlab{a}})\citenamefont {Wang}, \citenamefont {Huang},
  \citenamefont {Zhou}, \citenamefont {Liu}, \citenamefont {Ma}, \citenamefont
  {Wang},\ and\ \citenamefont {Zeng}}]{wang2018high}%
  \BibitemOpen
  \bibfield  {author} {\bibinfo {author} {\bibfnamefont {T.}~\bibnamefont
  {Wang}}, \bibinfo {author} {\bibfnamefont {P.}~\bibnamefont {Huang}},
  \bibinfo {author} {\bibfnamefont {Y.}~\bibnamefont {Zhou}}, \bibinfo {author}
  {\bibfnamefont {W.}~\bibnamefont {Liu}}, \bibinfo {author} {\bibfnamefont
  {H.}~\bibnamefont {Ma}}, \bibinfo {author} {\bibfnamefont {S.}~\bibnamefont
  {Wang}},\ and\ \bibinfo {author} {\bibfnamefont {G.}~\bibnamefont {Zeng}},\
  }\bibfield  {title} {\bibinfo {title} {High key rate continuous-variable
  quantum key distribution with a real local oscillator},\ }\href
  {https://doi.org/10.1364/OE.26.002794} {\bibfield  {journal} {\bibinfo
  {journal} {Opt. Express}\ }\textbf {\bibinfo {volume} {26}},\ \bibinfo
  {pages} {2794} (\bibinfo {year} {2018}{\natexlab{a}})}\BibitemShut {NoStop}%
\bibitem [{\citenamefont {Huang}\ \emph {et~al.}(2016)\citenamefont {Huang},
  \citenamefont {Huang}, \citenamefont {Lin},\ and\ \citenamefont
  {Zeng}}]{huang2016long}%
  \BibitemOpen
  \bibfield  {author} {\bibinfo {author} {\bibfnamefont {D.}~\bibnamefont
  {Huang}}, \bibinfo {author} {\bibfnamefont {P.}~\bibnamefont {Huang}},
  \bibinfo {author} {\bibfnamefont {D.}~\bibnamefont {Lin}},\ and\ \bibinfo
  {author} {\bibfnamefont {G.}~\bibnamefont {Zeng}},\ }\bibfield  {title}
  {\bibinfo {title} {Long-distance continuous-variable quantum key distribution
  by controlling excess noise},\ }\href
  {https://www.nature.com/articles/srep19201} {\bibfield  {journal} {\bibinfo
  {journal} {Scientific reports}\ }\textbf {\bibinfo {volume} {6}},\ \bibinfo
  {pages} {19201} (\bibinfo {year} {2016})}\BibitemShut {NoStop}%
\bibitem [{\citenamefont {Wang}\ \emph
  {et~al.}(2018{\natexlab{b}})\citenamefont {Wang}, \citenamefont {Du},
  \citenamefont {Liu}, \citenamefont {Wang}, \citenamefont {Li},\ and\
  \citenamefont {Peng}}]{wang2018long}%
  \BibitemOpen
  \bibfield  {author} {\bibinfo {author} {\bibfnamefont {N.}~\bibnamefont
  {Wang}}, \bibinfo {author} {\bibfnamefont {S.}~\bibnamefont {Du}}, \bibinfo
  {author} {\bibfnamefont {W.}~\bibnamefont {Liu}}, \bibinfo {author}
  {\bibfnamefont {X.}~\bibnamefont {Wang}}, \bibinfo {author} {\bibfnamefont
  {Y.}~\bibnamefont {Li}},\ and\ \bibinfo {author} {\bibfnamefont
  {K.}~\bibnamefont {Peng}},\ }\bibfield  {title} {\bibinfo {title}
  {Long-distance continuous-variable quantum key distribution with entangled
  states},\ }\href {https://doi.org/10.1103/PhysRevApplied.10.064028}
  {\bibfield  {journal} {\bibinfo  {journal} {Phys. Rev. Appl.}\ }\textbf
  {\bibinfo {volume} {10}},\ \bibinfo {pages} {064028} (\bibinfo {year}
  {2018}{\natexlab{b}})}\BibitemShut {NoStop}%
\bibitem [{\citenamefont {Ghorai}\ \emph {et~al.}(2019)\citenamefont {Ghorai},
  \citenamefont {Grangier}, \citenamefont {Diamanti},\ and\ \citenamefont
  {Leverrier}}]{ghorai2019asymptotic}%
  \BibitemOpen
  \bibfield  {author} {\bibinfo {author} {\bibfnamefont {S.}~\bibnamefont
  {Ghorai}}, \bibinfo {author} {\bibfnamefont {P.}~\bibnamefont {Grangier}},
  \bibinfo {author} {\bibfnamefont {E.}~\bibnamefont {Diamanti}},\ and\
  \bibinfo {author} {\bibfnamefont {A.}~\bibnamefont {Leverrier}},\ }\bibfield
  {title} {\bibinfo {title} {Asymptotic security of continuous-variable quantum
  key distribution with a discrete modulation},\ }\href
  {https://doi.org/10.1103/PhysRevX.9.021059} {\bibfield  {journal} {\bibinfo
  {journal} {Phys. Rev. X}\ }\textbf {\bibinfo {volume} {9}},\ \bibinfo {pages}
  {021059} (\bibinfo {year} {2019})}\BibitemShut {NoStop}%
\bibitem [{\citenamefont {Lin}\ \emph {et~al.}(2019)\citenamefont {Lin},
  \citenamefont {Upadhyaya},\ and\ \citenamefont
  {L\"utkenhaus}}]{lin2019asymptotic}%
  \BibitemOpen
  \bibfield  {author} {\bibinfo {author} {\bibfnamefont {J.}~\bibnamefont
  {Lin}}, \bibinfo {author} {\bibfnamefont {T.}~\bibnamefont {Upadhyaya}},\
  and\ \bibinfo {author} {\bibfnamefont {N.}~\bibnamefont {L\"utkenhaus}},\
  }\bibfield  {title} {\bibinfo {title} {Asymptotic security analysis of
  discrete-modulated continuous-variable quantum key distribution},\ }\href
  {https://doi.org/10.1103/PhysRevX.9.041064} {\bibfield  {journal} {\bibinfo
  {journal} {Phys. Rev. X}\ }\textbf {\bibinfo {volume} {9}},\ \bibinfo {pages}
  {041064} (\bibinfo {year} {2019})}\BibitemShut {NoStop}%
\bibitem [{\citenamefont {Huang}\ \emph {et~al.}(2021)\citenamefont {Huang},
  \citenamefont {Shen}, \citenamefont {Wang}, \citenamefont {Chen},
  \citenamefont {Xu}, \citenamefont {Yu},\ and\ \citenamefont
  {Guo}}]{huang2021realizing}%
  \BibitemOpen
  \bibfield  {author} {\bibinfo {author} {\bibfnamefont {Y.}~\bibnamefont
  {Huang}}, \bibinfo {author} {\bibfnamefont {T.}~\bibnamefont {Shen}},
  \bibinfo {author} {\bibfnamefont {X.}~\bibnamefont {Wang}}, \bibinfo {author}
  {\bibfnamefont {Z.}~\bibnamefont {Chen}}, \bibinfo {author} {\bibfnamefont
  {B.}~\bibnamefont {Xu}}, \bibinfo {author} {\bibfnamefont {S.}~\bibnamefont
  {Yu}},\ and\ \bibinfo {author} {\bibfnamefont {H.}~\bibnamefont {Guo}},\
  }\bibfield  {title} {\bibinfo {title} {Realizing a downstream-access network
  using continuous-variable quantum key distribution},\ }\href
  {https://doi.org/10.1103/PhysRevApplied.16.064051} {\bibfield  {journal}
  {\bibinfo  {journal} {Phys. Rev. Appl.}\ }\textbf {\bibinfo {volume} {16}},\
  \bibinfo {pages} {064051} (\bibinfo {year} {2021})}\BibitemShut {NoStop}%
\bibitem [{\citenamefont {Tian}\ \emph {et~al.}(2022)\citenamefont {Tian},
  \citenamefont {Wang}, \citenamefont {Liu}, \citenamefont {Du}, \citenamefont
  {Liu}, \citenamefont {Lu}, \citenamefont {Wang},\ and\ \citenamefont
  {Li}}]{tian2022experimental}%
  \BibitemOpen
  \bibfield  {author} {\bibinfo {author} {\bibfnamefont {Y.}~\bibnamefont
  {Tian}}, \bibinfo {author} {\bibfnamefont {P.}~\bibnamefont {Wang}}, \bibinfo
  {author} {\bibfnamefont {J.}~\bibnamefont {Liu}}, \bibinfo {author}
  {\bibfnamefont {S.}~\bibnamefont {Du}}, \bibinfo {author} {\bibfnamefont
  {W.}~\bibnamefont {Liu}}, \bibinfo {author} {\bibfnamefont {Z.}~\bibnamefont
  {Lu}}, \bibinfo {author} {\bibfnamefont {X.}~\bibnamefont {Wang}},\ and\
  \bibinfo {author} {\bibfnamefont {Y.}~\bibnamefont {Li}},\ }\bibfield
  {title} {\bibinfo {title} {Experimental demonstration of continuous-variable
  measurement-device-independent quantum key distribution over optical fiber},\
  }\href@noop {} {\bibfield  {journal} {\bibinfo  {journal} {Optica}\ }\textbf
  {\bibinfo {volume} {9}},\ \bibinfo {pages} {492} (\bibinfo {year}
  {2022})}\BibitemShut {NoStop}%
\bibitem [{\citenamefont {Wang}\ \emph
  {et~al.}(2022{\natexlab{b}})\citenamefont {Wang}, \citenamefont {Li},
  \citenamefont {Pi}, \citenamefont {Pan}, \citenamefont {Shao}, \citenamefont
  {Ma}, \citenamefont {Zhang}, \citenamefont {Yang}, \citenamefont {Zhang},
  \citenamefont {Huang} \emph {et~al.}}]{wang2022sub}%
  \BibitemOpen
  \bibfield  {author} {\bibinfo {author} {\bibfnamefont {H.}~\bibnamefont
  {Wang}}, \bibinfo {author} {\bibfnamefont {Y.}~\bibnamefont {Li}}, \bibinfo
  {author} {\bibfnamefont {Y.}~\bibnamefont {Pi}}, \bibinfo {author}
  {\bibfnamefont {Y.}~\bibnamefont {Pan}}, \bibinfo {author} {\bibfnamefont
  {Y.}~\bibnamefont {Shao}}, \bibinfo {author} {\bibfnamefont {L.}~\bibnamefont
  {Ma}}, \bibinfo {author} {\bibfnamefont {Y.}~\bibnamefont {Zhang}}, \bibinfo
  {author} {\bibfnamefont {J.}~\bibnamefont {Yang}}, \bibinfo {author}
  {\bibfnamefont {T.}~\bibnamefont {Zhang}}, \bibinfo {author} {\bibfnamefont
  {W.}~\bibnamefont {Huang}}, \emph {et~al.},\ }\bibfield  {title} {\bibinfo
  {title} {Sub-gbps key rate four-state continuous-variable quantum key
  distribution within metropolitan area},\ }\href
  {https://www.nature.com/articles/s42005-022-00941-z} {\bibfield  {journal}
  {\bibinfo  {journal} {Communications Physics}\ }\textbf {\bibinfo {volume}
  {5}},\ \bibinfo {pages} {162} (\bibinfo {year}
  {2022}{\natexlab{b}})}\BibitemShut {NoStop}%
\bibitem [{\citenamefont {Zhou}\ \emph {et~al.}(2019)\citenamefont {Zhou},
  \citenamefont {Wang}, \citenamefont {Zhang}, \citenamefont {Zhang},
  \citenamefont {Yu},\ and\ \citenamefont {Guo}}]{zhou2019continuous}%
  \BibitemOpen
  \bibfield  {author} {\bibinfo {author} {\bibfnamefont {C.}~\bibnamefont
  {Zhou}}, \bibinfo {author} {\bibfnamefont {X.}~\bibnamefont {Wang}}, \bibinfo
  {author} {\bibfnamefont {Y.}~\bibnamefont {Zhang}}, \bibinfo {author}
  {\bibfnamefont {Z.}~\bibnamefont {Zhang}}, \bibinfo {author} {\bibfnamefont
  {S.}~\bibnamefont {Yu}},\ and\ \bibinfo {author} {\bibfnamefont
  {H.}~\bibnamefont {Guo}},\ }\bibfield  {title} {\bibinfo {title}
  {Continuous-variable quantum key distribution with rateless reconciliation
  protocol},\ }\href {https://doi.org/10.1103/PhysRevApplied.12.054013}
  {\bibfield  {journal} {\bibinfo  {journal} {Phys. Rev. Appl.}\ }\textbf
  {\bibinfo {volume} {12}},\ \bibinfo {pages} {054013} (\bibinfo {year}
  {2019})}\BibitemShut {NoStop}%
\bibitem [{\citenamefont {Blandino}\ \emph {et~al.}(2012)\citenamefont
  {Blandino}, \citenamefont {Leverrier}, \citenamefont {Barbieri},
  \citenamefont {Etesse}, \citenamefont {Grangier},\ and\ \citenamefont
  {Tualle-Brouri}}]{blandino2012improving}%
  \BibitemOpen
  \bibfield  {author} {\bibinfo {author} {\bibfnamefont {R.}~\bibnamefont
  {Blandino}}, \bibinfo {author} {\bibfnamefont {A.}~\bibnamefont {Leverrier}},
  \bibinfo {author} {\bibfnamefont {M.}~\bibnamefont {Barbieri}}, \bibinfo
  {author} {\bibfnamefont {J.}~\bibnamefont {Etesse}}, \bibinfo {author}
  {\bibfnamefont {P.}~\bibnamefont {Grangier}},\ and\ \bibinfo {author}
  {\bibfnamefont {R.}~\bibnamefont {Tualle-Brouri}},\ }\bibfield  {title}
  {\bibinfo {title} {Improving the maximum transmission distance of
  continuous-variable quantum key distribution using a noiseless amplifier},\
  }\href {https://doi.org/10.1103/PhysRevA.86.012327} {\bibfield  {journal}
  {\bibinfo  {journal} {Phys. Rev. A}\ }\textbf {\bibinfo {volume} {86}},\
  \bibinfo {pages} {012327} (\bibinfo {year} {2012})}\BibitemShut {NoStop}%
\bibitem [{\citenamefont {Opatrn\'y}\ \emph {et~al.}(2000)\citenamefont
  {Opatrn\'y}, \citenamefont {Kurizki},\ and\ \citenamefont
  {Welsch}}]{opatrny2000improvement}%
  \BibitemOpen
  \bibfield  {author} {\bibinfo {author} {\bibfnamefont {T.}~\bibnamefont
  {Opatrn\'y}}, \bibinfo {author} {\bibfnamefont {G.}~\bibnamefont {Kurizki}},\
  and\ \bibinfo {author} {\bibfnamefont {D.-G.}\ \bibnamefont {Welsch}},\
  }\bibfield  {title} {\bibinfo {title} {Improvement on teleportation of
  continuous variables by photon subtraction via conditional measurement},\
  }\href {https://doi.org/10.1103/PhysRevA.61.032302} {\bibfield  {journal}
  {\bibinfo  {journal} {Phys. Rev. A}\ }\textbf {\bibinfo {volume} {61}},\
  \bibinfo {pages} {032302} (\bibinfo {year} {2000})}\BibitemShut {NoStop}%
\bibitem [{\citenamefont {Huang}\ \emph {et~al.}(2013)\citenamefont {Huang},
  \citenamefont {He}, \citenamefont {Fang},\ and\ \citenamefont
  {Zeng}}]{huang2013performance}%
  \BibitemOpen
  \bibfield  {author} {\bibinfo {author} {\bibfnamefont {P.}~\bibnamefont
  {Huang}}, \bibinfo {author} {\bibfnamefont {G.}~\bibnamefont {He}}, \bibinfo
  {author} {\bibfnamefont {J.}~\bibnamefont {Fang}},\ and\ \bibinfo {author}
  {\bibfnamefont {G.}~\bibnamefont {Zeng}},\ }\bibfield  {title} {\bibinfo
  {title} {Performance improvement of continuous-variable quantum key
  distribution via photon subtraction},\ }\href
  {https://doi.org/10.1103/PhysRevA.87.012317} {\bibfield  {journal} {\bibinfo
  {journal} {Phys. Rev. A}\ }\textbf {\bibinfo {volume} {87}},\ \bibinfo
  {pages} {012317} (\bibinfo {year} {2013})}\BibitemShut {NoStop}%
\bibitem [{\citenamefont {Fiur\'a\ifmmode~\check{s}\else \v{s}\fi{}ek}\ and\
  \citenamefont {Cerf}(2012)}]{fiuravsek2012gaussian}%
  \BibitemOpen
  \bibfield  {author} {\bibinfo {author} {\bibfnamefont {J.}~\bibnamefont
  {Fiur\'a\ifmmode~\check{s}\else \v{s}\fi{}ek}}\ and\ \bibinfo {author}
  {\bibfnamefont {N.~J.}\ \bibnamefont {Cerf}},\ }\bibfield  {title} {\bibinfo
  {title} {Gaussian postselection and virtual noiseless amplification in
  continuous-variable quantum key distribution},\ }\href
  {https://doi.org/10.1103/PhysRevA.86.060302} {\bibfield  {journal} {\bibinfo
  {journal} {Phys. Rev. A}\ }\textbf {\bibinfo {volume} {86}},\ \bibinfo
  {pages} {060302} (\bibinfo {year} {2012})}\BibitemShut {NoStop}%
\bibitem [{\citenamefont {Chrzanowski}\ \emph {et~al.}(2014)\citenamefont
  {Chrzanowski}, \citenamefont {Walk}, \citenamefont {Assad}, \citenamefont
  {Janousek}, \citenamefont {Hosseini}, \citenamefont {Ralph}, \citenamefont
  {Symul},\ and\ \citenamefont {Lam}}]{chrzanowski2014measurement}%
  \BibitemOpen
  \bibfield  {author} {\bibinfo {author} {\bibfnamefont {H.~M.}\ \bibnamefont
  {Chrzanowski}}, \bibinfo {author} {\bibfnamefont {N.}~\bibnamefont {Walk}},
  \bibinfo {author} {\bibfnamefont {S.~M.}\ \bibnamefont {Assad}}, \bibinfo
  {author} {\bibfnamefont {J.}~\bibnamefont {Janousek}}, \bibinfo {author}
  {\bibfnamefont {S.}~\bibnamefont {Hosseini}}, \bibinfo {author}
  {\bibfnamefont {T.~C.}\ \bibnamefont {Ralph}}, \bibinfo {author}
  {\bibfnamefont {T.}~\bibnamefont {Symul}},\ and\ \bibinfo {author}
  {\bibfnamefont {P.~K.}\ \bibnamefont {Lam}},\ }\bibfield  {title} {\bibinfo
  {title} {Measurement-based noiseless linear amplification for quantum
  communication},\ }\href {https://www.nature.com/articles/nphoton.2014.49}
  {\bibfield  {journal} {\bibinfo  {journal} {Nature Photonics}\ }\textbf
  {\bibinfo {volume} {8}},\ \bibinfo {pages} {333} (\bibinfo {year}
  {2014})}\BibitemShut {NoStop}%
\bibitem [{\citenamefont {Li}\ \emph {et~al.}(2016)\citenamefont {Li},
  \citenamefont {Zhang}, \citenamefont {Wang}, \citenamefont {Xu},
  \citenamefont {Peng},\ and\ \citenamefont {Guo}}]{li2016non}%
  \BibitemOpen
  \bibfield  {author} {\bibinfo {author} {\bibfnamefont {Z.}~\bibnamefont
  {Li}}, \bibinfo {author} {\bibfnamefont {Y.}~\bibnamefont {Zhang}}, \bibinfo
  {author} {\bibfnamefont {X.}~\bibnamefont {Wang}}, \bibinfo {author}
  {\bibfnamefont {B.}~\bibnamefont {Xu}}, \bibinfo {author} {\bibfnamefont
  {X.}~\bibnamefont {Peng}},\ and\ \bibinfo {author} {\bibfnamefont
  {H.}~\bibnamefont {Guo}},\ }\bibfield  {title} {\bibinfo {title}
  {Non-gaussian postselection and virtual photon subtraction in
  continuous-variable quantum key distribution},\ }\href
  {https://doi.org/10.1103/PhysRevA.93.012310} {\bibfield  {journal} {\bibinfo
  {journal} {Phys. Rev. A}\ }\textbf {\bibinfo {volume} {93}},\ \bibinfo
  {pages} {012310} (\bibinfo {year} {2016})}\BibitemShut {NoStop}%
\bibitem [{\citenamefont {Lodewyck}\ \emph {et~al.}(2007)\citenamefont
  {Lodewyck}, \citenamefont {Bloch}, \citenamefont {Garc\'{\i}a-Patr\'on},
  \citenamefont {Fossier}, \citenamefont {Karpov}, \citenamefont {Diamanti},
  \citenamefont {Debuisschert}, \citenamefont {Cerf}, \citenamefont
  {Tualle-Brouri}, \citenamefont {McLaughlin},\ and\ \citenamefont
  {Grangier}}]{lodewyck2007quantum}%
  \BibitemOpen
  \bibfield  {author} {\bibinfo {author} {\bibfnamefont {J.}~\bibnamefont
  {Lodewyck}}, \bibinfo {author} {\bibfnamefont {M.}~\bibnamefont {Bloch}},
  \bibinfo {author} {\bibfnamefont {R.}~\bibnamefont {Garc\'{\i}a-Patr\'on}},
  \bibinfo {author} {\bibfnamefont {S.}~\bibnamefont {Fossier}}, \bibinfo
  {author} {\bibfnamefont {E.}~\bibnamefont {Karpov}}, \bibinfo {author}
  {\bibfnamefont {E.}~\bibnamefont {Diamanti}}, \bibinfo {author}
  {\bibfnamefont {T.}~\bibnamefont {Debuisschert}}, \bibinfo {author}
  {\bibfnamefont {N.~J.}\ \bibnamefont {Cerf}}, \bibinfo {author}
  {\bibfnamefont {R.}~\bibnamefont {Tualle-Brouri}}, \bibinfo {author}
  {\bibfnamefont {S.~W.}\ \bibnamefont {McLaughlin}},\ and\ \bibinfo {author}
  {\bibfnamefont {P.}~\bibnamefont {Grangier}},\ }\bibfield  {title} {\bibinfo
  {title} {Quantum key distribution over
  $25\phantom{\rule{0.3em}{0ex}}\mathrm{km}$ with an all-fiber
  continuous-variable system},\ }\href
  {https://doi.org/10.1103/PhysRevA.76.042305} {\bibfield  {journal} {\bibinfo
  {journal} {Phys. Rev. A}\ }\textbf {\bibinfo {volume} {76}},\ \bibinfo
  {pages} {042305} (\bibinfo {year} {2007})}\BibitemShut {NoStop}%
\bibitem [{\citenamefont {Leverrier}\ \emph {et~al.}(2008)\citenamefont
  {Leverrier}, \citenamefont {All\'eaume}, \citenamefont {Boutros},
  \citenamefont {Z\'emor},\ and\ \citenamefont
  {Grangier}}]{leverrier2008multidimensional}%
  \BibitemOpen
  \bibfield  {author} {\bibinfo {author} {\bibfnamefont {A.}~\bibnamefont
  {Leverrier}}, \bibinfo {author} {\bibfnamefont {R.}~\bibnamefont
  {All\'eaume}}, \bibinfo {author} {\bibfnamefont {J.}~\bibnamefont {Boutros}},
  \bibinfo {author} {\bibfnamefont {G.}~\bibnamefont {Z\'emor}},\ and\ \bibinfo
  {author} {\bibfnamefont {P.}~\bibnamefont {Grangier}},\ }\bibfield  {title}
  {\bibinfo {title} {Multidimensional reconciliation for a continuous-variable
  quantum key distribution},\ }\href
  {https://doi.org/10.1103/PhysRevA.77.042325} {\bibfield  {journal} {\bibinfo
  {journal} {Phys. Rev. A}\ }\textbf {\bibinfo {volume} {77}},\ \bibinfo
  {pages} {042325} (\bibinfo {year} {2008})}\BibitemShut {NoStop}%
\bibitem [{\citenamefont {Shokrollahi}(2006)}]{shokrollahi2006raptor}%
  \BibitemOpen
  \bibfield  {author} {\bibinfo {author} {\bibfnamefont {A.}~\bibnamefont
  {Shokrollahi}},\ }\bibfield  {title} {\bibinfo {title} {Raptor codes},\
  }\href {https://doi.org/10.1109/TIT.2006.874390} {\bibfield  {journal}
  {\bibinfo  {journal} {IEEE Transactions on Information Theory}\ }\textbf
  {\bibinfo {volume} {52}},\ \bibinfo {pages} {2551} (\bibinfo {year}
  {2006})}\BibitemShut {NoStop}%
\bibitem [{\citenamefont {Arikan}(2009)}]{arikan2009channel}%
  \BibitemOpen
  \bibfield  {author} {\bibinfo {author} {\bibfnamefont {E.}~\bibnamefont
  {Arikan}},\ }\bibfield  {title} {\bibinfo {title} {Channel polarization: A
  method for constructing capacity-achieving codes for symmetric binary-input
  memoryless channels},\ }\href {https://doi.org/10.1109/TIT.2009.2021379}
  {\bibfield  {journal} {\bibinfo  {journal} {IEEE Transactions on Information
  Theory}\ }\textbf {\bibinfo {volume} {55}},\ \bibinfo {pages} {3051}
  (\bibinfo {year} {2009})}\BibitemShut {NoStop}%
\bibitem [{\citenamefont {Richardson}\ and\ \citenamefont
  {Urbanke}(2008)}]{richardson2008modern}%
  \BibitemOpen
  \bibfield  {author} {\bibinfo {author} {\bibfnamefont {T.}~\bibnamefont
  {Richardson}}\ and\ \bibinfo {author} {\bibfnamefont {R.}~\bibnamefont
  {Urbanke}},\ }\href@noop {} {\emph {\bibinfo {title} {Modern coding
  theory}}}\ (\bibinfo  {publisher} {Cambridge university press},\ \bibinfo
  {year} {2008})\BibitemShut {NoStop}%
\bibitem [{\citenamefont {Hocevar}(2004)}]{hocevar2004reduced}%
  \BibitemOpen
  \bibfield  {author} {\bibinfo {author} {\bibfnamefont {D.}~\bibnamefont
  {Hocevar}},\ }\bibfield  {title} {\bibinfo {title} {A reduced complexity
  decoder architecture via layered decoding of ldpc codes},\ }in\ \href
  {https://doi.org/10.1109/SIPS.2004.1363033} {\emph {\bibinfo {booktitle}
  {IEEE Workshop onSignal Processing Systems, 2004. SIPS 2004.}}}\ (\bibinfo
  {year} {2004})\ pp.\ \bibinfo {pages} {107--112}\BibitemShut {NoStop}%
\bibitem [{\citenamefont {Navascu\'es}\ \emph {et~al.}(2006)\citenamefont
  {Navascu\'es}, \citenamefont {Grosshans},\ and\ \citenamefont
  {Ac\'{\i}n}}]{41}%
  \BibitemOpen
  \bibfield  {author} {\bibinfo {author} {\bibfnamefont {M.}~\bibnamefont
  {Navascu\'es}}, \bibinfo {author} {\bibfnamefont {F.}~\bibnamefont
  {Grosshans}},\ and\ \bibinfo {author} {\bibfnamefont {A.}~\bibnamefont
  {Ac\'{\i}n}},\ }\bibfield  {title} {\bibinfo {title} {Optimality of gaussian
  attacks in continuous-variable quantum cryptography},\ }\href
  {https://doi.org/10.1103/PhysRevLett.97.190502} {\bibfield  {journal}
  {\bibinfo  {journal} {Phys. Rev. Lett.}\ }\textbf {\bibinfo {volume} {97}},\
  \bibinfo {pages} {190502} (\bibinfo {year} {2006})}\BibitemShut {NoStop}%
\bibitem [{\citenamefont {Wolf}\ \emph {et~al.}(2006)\citenamefont {Wolf},
  \citenamefont {Giedke},\ and\ \citenamefont {Cirac}}]{42}%
  \BibitemOpen
  \bibfield  {author} {\bibinfo {author} {\bibfnamefont {M.~M.}\ \bibnamefont
  {Wolf}}, \bibinfo {author} {\bibfnamefont {G.}~\bibnamefont {Giedke}},\ and\
  \bibinfo {author} {\bibfnamefont {J.~I.}\ \bibnamefont {Cirac}},\ }\bibfield
  {title} {\bibinfo {title} {Extremality of gaussian quantum states},\ }\href
  {https://doi.org/10.1103/PhysRevLett.96.080502} {\bibfield  {journal}
  {\bibinfo  {journal} {Phys. Rev. Lett.}\ }\textbf {\bibinfo {volume} {96}},\
  \bibinfo {pages} {080502} (\bibinfo {year} {2006})}\BibitemShut {NoStop}%
\end{thebibliography}%

\end{document}